\documentclass[a4paper,11pt]{article}
\pdfoutput=1 

\usepackage{jcappub} 

\DeclareMathOperator{\Tr}{tr}

\usepackage{placeins}
\usepackage[T1]{fontenc} 
\usepackage{graphicx} 
\usepackage{txfonts}
\usepackage{hyperref}
\usepackage{gensymb}
\usepackage{bm}
\usepackage{xcolor}
\usepackage{booktabs}
\usepackage{amsmath}

\title{Non-Gaussianity in SMICA}

\author[a,b,1]{M. Citran,\note{Corresponding author.}}
\author[a]{H.V. Tran,}
\author[c,d,e]{G. Patanchon,}
\author[b]{B. van Tent}

\affiliation[a]{Universit\'e Paris Cit\'e, CNRS, Astroparticule et Cosmologie, 75013 Paris, France}
\affiliation[b]{Universit\'e Paris-Saclay, CNRS/IN2P3, IJCLab, 91405 Orsay, France}
\affiliation[c]{ILANCE, CNRS – University of Tokyo International Research Laboratory, Kashiwa, Chiba
277-8582, Japan}
\affiliation[d]{Universit\'e Paris Cit\'e, F-75006 Paris, France}
\affiliation[e]{Kavli Institute for the Physics and Mathematics of the Universe (Kavli IPMU, WPI), UTIAS, The University of Tokyo, Kashiwa, Chiba 277-8583, Japan}

\emailAdd{citran@apc.in2p3.fr}
\emailAdd{tran@apc.in2p3.fr}
\emailAdd{Guillaume.Patanchon@apc.univ-paris-diderot.fr}
\emailAdd{bartjan.van-tent@ijclab.in2p3.fr}

\abstract{We develop a new formalism for the component separation method Spectral Matching Independent Component Analysis (SMICA) in order to include the information contained in the foregrounds beyond second-order statistics. We also develop a binned bispectrum estimator that works directly using maps of different frequency channels, capable of determining the bispectrum of multiple components at the same time, shifting the traditional approach to non-Gaussianity estimation from a cleaned map to the component separation step, for a better handling of foreground uncertainty. We test our method on 400 E and B polarization simulations based on the LiteBIRD experiment, containing the two main sources of contamination for CMB polarization experiments: polarized dust and synchrotron emission. We show that the bispectrum does not improve the precision of the power spectrum estimation or of the spectral parameters. However, we are capable of recovering the correct $3$-point correlator of the foregrounds and standard constraints on primordial non-Gaussianity  in a coherent multi-frequency and multi-component framework. The advantage of our approach is that it combines data in an optimal way accounting for the power spectrum and the bispectrum of the various components, which is not true for the standard approach.}

\begin{document}
\maketitle
\flushbottom

\section{Introduction}
\label{sec:introduction}
The Cosmic Microwave Background (CMB) is one of the most important sources of information regarding our Universe. Its statistical properties are related to those of the inflaton field(s) and therefore hold a strong constraining power over inflationary models and over the current cosmological standard model: the $\Lambda \text{CDM}$ model. In this regard, the $2$-point correlator, the power spectrum, of the distribution of the CMB fluctuations is the most important observable, since the CMB field is well approximated by a gaussian statistically isotropic field that contains valuable information on the cosmological parameters~\cite{PR3_1,PR3_2}. However, due to the non-linearity of the Einstein equations, any model of inflation is expected to produce a small amount of non-Gaussianity, with a lower bound given by single-field slow-roll inflation~\cite{Maldacena_2003}. Indeed, if measured, the size and type of non-Gaussianity in the CMB anisotropies will help us distinguish between single- and multi-field inflationary models~\cite{Vernizzi_2006, Rigopoulos_2007, Byrnes_2010, Jung_2017, de_Putter_2017, Wang_2023} or models with non-standard kinetic terms~\cite{Seery_2005, Chen_2007, Bartolo_2010}. Primordial Non-Gaussianity (PNG) has been the focus of 3 \textit{Planck} collaboration papers~\cite{Planck_2013, Planck_2015, Planck_2018} and continues to be central for current and  future cosmological experiments. Indeed, both future CMB missions, such as the Simons Observatory~\cite{Simons_2019} and the LiteBIRD satellite~\cite{LB_2025}, and the new generation of galaxy surveys capable of probing the three-dimensional information of the large-scale structure (LSS)---including Euclid~\cite{Euclid_2011}, SPHEREx~\cite{Spherex_2020}, and the Rubin Observatory~\cite{Rubin_2024}---will be able to significantly improve upon the current constraints obtained with the CMB.

The CMB is also polarized at the level of 10\% due to Thomson scattering: if an electromagnetic wave scatters off an electron, it creates linear polarization perpendicular to both the incoming and outgoing wave vectors. If the incoming radiation is isotropic, then a second, perpendicular, incoming wave will provide the other linear polarization direction, so that the outgoing wave is again unpolarized. However, if the incoming wave has a quadrupole moment, then the two linear polarization directions will not have the same amplitude, so that a net linear polarization remains for the outgoing wave. Instead of the usual Q and U Stokes parameters to describe linear polarization, in CMB physics it is common to use the E- and B-mode polarization. These are invariant under translations and rotations and are respectively a scalar (E) and a pseudo-scalar (B) field, but note that these are not the electric and magnetic fields. As scalar perturbations can only create E-polarization, not B, the primordial B-polarization signal will be a clear probe of the inflationary tensor perturbations. The tensor-to-scalar ratio $r$ quantifies the relative strength of the primordial gravitational waves compared to scalar fluctuations, and is defined as the ratio of the amplitudes of the primordial tensor and scalar power spectra
\begin{align}
    \label{first}
    r = \frac{A_t}{A_s}
\end{align}
Indeed, one of the main goals for future CMB missions, like the Simons Observatory~\cite{Simons_2019} and the LiteBIRD satellite~\cite{LB_2025}, will be to measure $r$ and provide the cosmological community with CMB B-mode polarization maps. In this work, we will study the LiteBIRD configuration: \textit{the Lite (Light) satellite for the study of B-mode polarization and Inflation from cosmic background Radiation Detection} is a space mission which aims to launch in the 2030's. Its current configuration contains 15 frequency channels in the range 40-400 GHz. One of the main challenges it will face is to distinguish the primordial B-mode CMB signal from galactic foreground emission, like thermal dust and synchrotron emission, which contaminate CMB measurements both at the level of the power spectrum and the bispectrum. For B-mode polarization, the signal of the foregrounds is significantly larger than the primordial one over the whole range of frequencies, making component separation a fundamental step to measure the primordial CMB signal. Moreover, the gravitational lensing of the primordial CMB E polarization can generate non-primordial B polarization, complicating the analysis. Many different component separation methods~\cite{Leloup_2023,Morshed_2024,Rizzieri_2025,Carones_2023, Vacher_2025} are being improved and developed for the new generation of CMB experiments.    

The standard analysis of CMB data can be schematized by independent steps consisting of: map making from time-ordered-data (TOD) coming from detectors, component separation from frequency maps, and finally CMB map analysis (power spectrum, bispectrum, trispectrum, etc.). Each of these steps works with different types of data (TODs, frequency maps, single CMB map), hence it employs a different set of assumptions concerning the data. The standard non-gaussian analysis of the CMB takes place at the end of this pipeline, while the CMB has usually been assumed to be fully gaussian during the previous component-separation step. With this work we aim to prevent this by advancing the analysis of the $3$-point correlator, the bispectrum, directly to the component separation step, working directly with frequency maps and different components. We adopt the framework of SMICA (Spectral Matching Independent Component Analysis)~\cite{Delabrouille_2003, Cardoso_2008}, a non-parametric component separation method that works in the harmonic domain, and we explore the possibility of including the non-Gaussianity of the signals in the component analysis. Indeed, the foreground components are not gaussian fields and we decided to make use of this information directly in the component separation step. For this purpose, we expand the distribution describing the CMB anisotropies around a gaussian distribution via the Multivariate Edgeworth Expansion (MEE) \cite{Juszkiewicz_1995, Amendola_1996, Taylor_2001, Bartolo_2012,  Hall_2022} truncated at the first order, allowing us to perform a combined analysis of the power spectrum and the bispectrum. Using a different approach from the MEE, as we found that it is not suitable for a bispectrum estimation due to having a linear dependence on the bispectrum, we develop a multi-component multi-frequency bispectrum estimator capable of determining the bispectrum of the foregrounds and of obtaining constraints on the PNG of the CMB that are compatible with the standard non-gaussian analysis \cite{Bucher_2010, Bucher_2016}, but directly in the component separation step and in a coherent simultaneous way for all the components.

We start by introducing the definition of harmonic space non-Gaussianity in section~\ref{sec:nonG}, defining the power spectrum and the bispectrum. We then discuss the existing framework of gaussian SMICA in section~\ref{sec:framework} and describe the simulations of LiteBIRD observations in section~\ref{sec:sim}. In section~\ref{Power_estimation} we introduce the new non-gaussian extension of SMICA: in section~\ref{sec:MEE} we describe the MEE, generalize it to multiple frequency channels, and write the likelihood we have used. Then, in section~\ref{subsec:datamodel} we introduce the bispectrum model we have implemented, as it was not possible to perform a joint estimation due to the already mentioned properties of the MEE, and in section~\ref{sec:res} we discuss the results of the power spectrum estimation conditioned by the bispectrum model. We then move on to the multi-frequency multi-component binned bispectrum estimator in section~\ref{sec:mdmcbb}: we introduce the new approach and likelihood we have implemented for the bispectrum estimation in section~\ref{sec:mdmcbb:lik}, then we define the binned bispectrum and its properties in section~\ref{sec:mdmcbb:bin} and finally we show the results we have obtained for the B-mode polarization and the E-mode polarization in section~\ref{sec:res2:B} and section~\ref{sec:res2:E} respectively. In the conclusion, section~\ref{sec:conc}, we summarize the results we have obtained and possible future developments. Appendix~\ref{app:MEE} contains the computations we performed for the MEE introduced in section~\ref{sec:MEE}.

\section{Non-Gaussianity}
\label{sec:nonG}
Non-Gaussianity is a general concept as it includes any deviation from a perfect gaussian distribution. In the case of CMB data analysis, we mainly work with random spherical fields $\Phi(\theta,\phi)$, defined on a sphere and decomposable into spherical harmonics $\mathcal{Y}_{lm}(\theta,\phi)$ as
\begin{align}
    \Phi(\theta,\phi) = \sum_{l=0}^{+\infty}\sum_{m=-l}^{+l}a_{lm}\mathcal{Y}_{lm}(\theta,\phi).
\end{align}
It is therefore possible to move to the harmonic space in order to deal with a discrete set of decorrelated random variables. For a statistically isotropic gaussian random spherical field~\cite{Marinucci_2011,Lang_2015}, the distribution of the spherical harmonic coefficients is gaussian with variance $C_l$, the power spectrum of the distribution defined by

\begin{align}
\label{eq:nonG:2}
    \langle a^*_{lm} a_{l'm'}\rangle = \delta_{ll'}\delta_{mm'}C_l ,
\end{align}
where the average is computed over an infinite set of skies. Of course in CMB data analysis we have access to only one set of $\widehat{a}_{lm}$ measured on the sky, hence it is important to distinguish the distribution's power spectrum from the observed one, which is computed by averaging modes with the same theoretical power spectrum
\begin{align}
\label{eq:nonG:3}
    \widehat{C}_l = \frac{1}{2l+1}\sum_{m=-l}^l |\widehat{a}_{lm}|^2.
\end{align}
This estimation is limited even in the case of perfect sky reconstruction by the number of modes per $l$ which is $2l+1$. This is what we call cosmic variance.

There are two ways to deviate from this distribution: anisotropy and non-Gaussianity. The first one can appear even in the gaussian case and  can be defined by a modified power spectrum
\begin{align}
\label{eq:nonG:4}
     \langle a^*_{lm} a_{l'm'}\rangle = C_{ll',mm'} ,
\end{align}
but in this work, we assume statistically isotropic fields. Instead, non-Gaussianity  appears when the distribution is no longer completely defined by its variance but has non-zero connected higher-order correlation functions. The first order is the 3-point correlator, called bispectrum, which in the case of a statistically isotropic field is given by
\begin{align}
\label{eq:nonG:4}
     \langle a_{l_1m_1} a_{l_2m_2} a_{l_3m_3}\rangle = \begin{pmatrix}
         l_1 & l_2 & l_3 \\ m_1 & m_2 & m_3
     \end{pmatrix} B_{l_1l_2l_3},
\end{align}
where $\begin{pmatrix} l_1 & l_2 & l_3 \\ m_1 & m_2 & m_3 \end{pmatrix}$ is the $3j$-symbol~\cite{Vars_1988}, whose properties we can use to invert eq.~\eqref{eq:nonG:4} and obtain
\begin{align}
\label{eq:nonG:6}
     B_{l_1l_2l_3} = \sum_{m_i}\begin{pmatrix}
         l_1 & l_2 & l_3 \\ m_1 & m_2 & m_3
     \end{pmatrix} \langle a_{l_1m_1} a_{l_2m_2} a_{l_3m_3}\rangle.
\end{align}
The observed bispectrum of a single sky is
\begin{align}
\label{eq:nonG:7}
     \widehat{B}_{l_1l_2l_3} = \sum_{m_i}\begin{pmatrix}
         l_1 & l_2 & l_3 \\ m_1 & m_2 & m_3
     \end{pmatrix}  \widehat{a}_{l_1m_1} \widehat{a}_{l_2m_2} \widehat{a}_{l_3m_3},
\end{align}
whose estimation is also limited by the number of modes present in each independent triplet $l_1\leq l_2\leq l_3$. Moreover, the bispectrum must essentially be divided into the even part with $l_1+l_2+l_3$ even and the odd part with $l_1+l_2+l_3$ odd. The first one is a real symmetric object while the second is purely imaginary and antisymmetric since
\begin{align}
\label{eq:nonG:8a}
     B_{l_1l_2l_3}^* =& \sum_{m_i}\begin{pmatrix}
         l_1 & l_2 & l_3 \\ m_1 & m_2 & m_3
     \end{pmatrix}  a_{l_1m_1}^* a_{l_2m_2}^* a_{l_3m_3}^* \nonumber\\
     =& \sum_{m_i}\begin{pmatrix}
         l_1 & l_2 & l_3 \\ m_1 & m_2 & m_3
     \end{pmatrix}  a_{l_1-m_1} a_{l_2-m_2} a_{l_3-m_3} \nonumber\\
     =& \sum_{m_i}\begin{pmatrix}
         l_1 & l_2 & l_3 \\ -m_1 & -m_2 & -m_3
     \end{pmatrix}  a_{l_1m_1} a_{l_2m_2} a_{l_3m_3} \nonumber\\
     =& (-1)^{l_1+l_2+l_3}\sum_{m_i}\begin{pmatrix}
         l_1 & l_2 & l_3 \\ m_1 & m_2 & m_3
     \end{pmatrix}  a_{l_1m_1} a_{l_2m_2} a_{l_3m_3} = (-1)^{l_1+l_2+l_3}B_{l_1l_2l_3}
\end{align}
and
\begin{align}
\label{eq:nonG:8b}
     B_{l_2l_1l_3} =& \sum_{m_i}\begin{pmatrix}
         l_2 & l_1 & l_3 \\ m_2 & m_1 & m_3
     \end{pmatrix}  a_{l_2m_2} a_{l_1m_1} a_{l_3m_3} \nonumber\\
     =& (-1)^{l_1+l_2+l_3}\sum_{m_i}\begin{pmatrix}
         l_1 & l_2 & l_3 \\ m_1 & m_2 & m_3
     \end{pmatrix}  a_{l_1m_1} a_{l_2m_2} a_{l_3m_3} = (-1)^{l_1+l_2+l_3}B_{l_1l_2l_3}.
\end{align}
This will be of extreme importance when working with a binned bispectrum.

Due to the limited number of modes in the sky, the observed bispectrum of one realization of a perfectly gaussian distribution is not zero; hence, the standard deviation of the bispectrum for a perfectly gaussian map represents the cosmic variance limit in the case of non-Gaussianity:
\begin{align}
\label{eq:nonG:9}
     \mathrm{Var}(\textbf{B}) = \langle B_{l_1l_2l_3}^2 \rangle - \langle B_{l_1l_2l_3} \rangle^2 = \langle B_{l_1l_2l_3}^2 \rangle  \propto C_{l_1}C_{l_2}C_{l_3} \neq 0.
\end{align}
Then, the limit of weak non-Gaussianity corresponds to
\begin{align}
\label{eq:nonG:10}
     \frac{B^2_{l_1l_2l_3}}{C_{l_1}C_{l_2}C_{l_3}}\ll 1,
\end{align}
which means a small bispectrum with respect to the gaussian standard deviation (and weak bispectrum signal-to-noise ratio). 
All of this can be generalized in the case of a sky observed at different frequencies $d$. We generalize the power spectrum via the covariance matrix $C^{dd'}_l$ defined by 
\begin{align}
\label{eq:nonG:11}
    \langle (a^d_{lm})^* a^{d'}_{l'm'} \rangle = \delta_{ll'}\delta_{mm'}C^{dd'}_{l},
\end{align}
and a bispectrum defined by
\begin{align}
\label{eq:nonG:12}
    \langle a^{d_1}_{l_1m_1} a^{d_2}_{l_2m_2} a^{d_3}_{l_3m_3}\rangle = \begin{pmatrix}
         l_1 & l_2 & l_3 \\ m_1 & m_2 & m_3
     \end{pmatrix} B^{d_1d_2d_3}_{l_1l_2l_3}.
\end{align}
Any other higher-order correlator (as the trispectrum~\cite{Philcox_2025}) is still a sign of non-Gaussianity, but 
in this work we restrict ourselves to the first-order correction to Gaussianity, i.e. the bispectrum.

Assuming that the cosmological signals, like the CMB, dust, synchrotron and other foregrounds, are random fields is the basis of many component separation methods. Assuming that the correlators are the same across the whole sky is often referred to as stationarity in component separation.  Statistical isotropy, which will be an assumption in the rest of this work, means that the correlators are rotationally invariant and implies stationarity.

\section{The multi-frequency multi-component spectral analysis framework}
\label{sec:framework}
The sky emission at millimeter wavelengths can be well modeled at first order by a linear combination of the emissions of different processes with constant emission spectra~\cite{Delabrouille_2003, Cardoso_2008}: CMB anisotropies, thermal dust emission, thermal Sunyaev Zel’dovich (SZ) effect, synchrotron emission etc., which we call components. Therefore, the observed sky at a given frequency channel $d$ is  a noisy linear mixture of $N_{\mathrm{comp}}$ components:

\begin{align}
\label{eq:frame:1}
    y^d(\theta,\phi)=& \sum_{c=1}^{N_{\mathrm{comp}}} A^{d c}s^c(\theta,\phi) + n^d(\theta,\phi),
\end{align}
where $s^c(\theta,\phi)$ represents the contribution of the $c$-th component while the coefficient $A^{dc}$ reflects the emission law and detector properties for the $c$-th component with respect to the $d$-th frequency and $n^d$ is the noise for the frequency channel $d$.

We can rewrite eq.~\eqref{eq:frame:1} in terms of spherical harmonics coefficients as
\begin{align}
\label{eq:frame:2}
    a^d_{lm}=& \sum_{c=1}^{N_{\mathrm{comp}}} A^{d c}s^c_{lm} + n^d_{lm} \ .
\end{align}
Now, under the assumptions of an isotropic and gaussian sky~\cite{Marinucci_2011,Lang_2015} we know that the distribution of the spherical harmonics coefficients $a^d_{lm}$ is a gaussian distribution completely described by the covariance matrix $C^{dd'}_l$ defined by eq.~\eqref{eq:nonG:11}, which for our linear sky model is
\begin{align}
\label{eq:frame:3}
   C^{dd'}_{l}=& \left\langle \left(\sum_{c=1}^{N_{\mathrm{comp}}} A^{d c}s^c_{lm} + n^d_{lm}\right)^*\left(\sum_{c'=1}^{N_{\mathrm{comp}}} A^{d' c'}s^{c'}_{lm} + n^{d'}_{lm}\right)\right\rangle \ .
\end{align}
Moreover, if we assume independence between components and uncorrelated noise across frequency channels, we obtain from eq.~\eqref{eq:frame:3}
\begin{align}
\label{eq:frame:4a}
   C^{dd'}_{l}=& \sum_c A^{d c} C^c_{l} A^{d' c} + (\sigma^d_{l})^2 
\end{align}
or in matrix notation
\begin{align}
\label{eq:frame:4b}
   \textbf{C}_l =& A\textbf{C}_s(l)A^T +\textbf{C}_n(l),\nonumber\\
   \textbf{C}_s(l)=&  \mathrm{diag}(C^1_l,...,C^{N_{\text{comp}}}_l),\nonumber\\
   \textbf{C}_n(l)=&  \mathrm{diag}((\sigma^1_{l})^2,...,(\sigma^{N_{\text{freq}}}_{l})^2),
\end{align}
where $C^c_l$ is the power-spectrum of the $c$-th component and $\sigma^d_{l}$ is the noise amplitude of the $d$-th frequency channel.

This implies that, given our linear model, the distribution of the spherical harmonic coefficients $a^d_{lm}$ is completely defined by the set of parameters (via the covariance matrix)
\begin{align}
\label{eq:frame:5}
   \vec{\theta} = \{ A^{dc}, C^c_l, \sigma^d_l\} \ .
\end{align}
Now, given one observed sky $\widehat{y}^d(\theta,\phi)$ over a set of frequencies and its spherical harmonics decomposition $\widehat{a}^d_{lm}$, we can write down the gaussian likelihood for our model
\begin{align}
\label{eq:frame:6}
   \mathcal{L}(\vec{\theta}\ |\ \widehat{a}^d_{lm}) = \prod_{lm} \frac{1}{\sqrt{2\pi \det(\textbf{C}(\vec{\theta})_l)}}\exp\left(-\frac{1}{2}(\widehat{a}^d_{lm})^*(C^{-1}(\vec{\theta}))^{dd'}_l\widehat{a}^d_{lm} \right)  .
\end{align}
It is simple to prove that this is equivalent to the following negative log-likelihood~\cite{Delabrouille_2003}
\begin{align}
\label{eq:frame:7}
   -2\log(\mathcal{L}(\vec{\theta}\ |\ \widehat{a}^d_{lm})) = \sum_l (2l+1)D(\widehat{\textbf{C}}_l,\textbf{C}(\vec{\theta})_l) + \mathrm{const} \ ,
\end{align}
where $D(\widehat{\textbf{C}}_l,\textbf{C}_l)$ is the so-called Kullback–Leibler divergence for $n \times n$ matrices
\begin{align}
\label{eq:frame:8}
   D(\textbf{R}_1,\textbf{R}_2) = \Tr(\textbf{R}_1\textbf{R}_2^{-1}) - \log(\det(\textbf{R}_1\textbf{R}_2^{-1}))
\end{align}
between the model covariance matrix and the observed covariance matrix defined as
\begin{align}
\label{eq:frame:9}
   \widehat{C}^{dd'}_l = \frac{1}{2l+1}\sum_m \widehat{a}^{d}_{lm}(\widehat{a}^{d'}_{lm})^* \ .
\end{align}

The SMICA (Spectral Matching Independent Component Analysis) method~\cite{Delabrouille_2003, Cardoso_2008} minimizes eq.~\eqref{eq:frame:7} to recover the spectral parameters $\vec{\theta}$ in eq.~\eqref{eq:frame:5}: the mixing matrix for each component, the power spectrum (for a given polarization mode) of each component and the noise per frequency channel. SMICA does not assume a parametric form for the mixing matrix $A$ but fits it blindly, which is allowed because of the diversity of the components' power spectra (non-proportional power spectra).

\section{Simulations}
\label{sec:sim}
For this project we work with 400 simulations based on the LiteBIRD configuration presented in~\cite{LB_2025}. One of the mission's science goals is measuring the B-mode CMB map, with unprecedented precision in order to constrain the tensor to scalar ratio $r$ and the energy scale at which inflation happened.

In order to be able to claim a detection of the primordial B-mode signal created by tensor fluctuations, the large scales of the Universe play a pivotal role. Indeed, the two main features that we look for in the CMB B-mode power spectrum are the "recombination peak" at $\ell \sim 80$, imprinted during the epoch when electrons and protons combine to form hydrogen and the Universe becomes neutral, and the "reionization bump" at $\ell\lesssim 10$,  imprinted around the time when the first stars reionize the Universe. At smaller scales the dominant effect is the weak gravitational lensing of the CMB E-modes which generates late time B-modes. For these reasons, and given the scaling of computational time with $\mathcal{O}(\ell^3)$, we restrict our analysis to $\ell_{\text{max}}=100$.

\begin{table}[tbp]
\centering
\setlength{\tabcolsep}{1pt}
\begin{tabular}{|c|c|c|c|c|c|c|c|c|c|c|c|c|c|c|c|}
\hline
$\nu$ [GHz]  & 40  & 50  & 60  & 68  & 78  & 89  & 100 & 119 & 140 & 166 & 195 & 235 & 280 & 337 & 402 \\
\hline
$\sigma_S$ [$\mu K\cdot \mathrm{arcmin}$] & 37.42 & 33.46 & 21.31 & 16.87 & 12.07 & 11.30 & 10.34 & 7.69  & 7.25  & 5.57  & 7.05  & 10.79 & 13.80 & 21.95 & 47.45\\
\hline
\end{tabular}
\caption{LiteBIRD's channel frequencies $\nu$ with the respective statistical sensitivities $\sigma_S$}
\label{tab:lb_rotated}
\end{table}

In table~\ref{tab:lb_rotated} we report LiteBIRD's channel frequencies $\nu$ with the respective statistical sensitivities $\sigma_S$ we used in our simulations. At these frequencies the observed B-mode field is well described by a combination of the CMB, thermal emission by interstellar dust grains aligned with the galactic magnetic field, synchrotron emission from accelerating electrons in the galactic magnetic field and, of course, instrumental noise. 

\begin{figure}[tbp]
\hspace*{-2cm} 
\centering 
\includegraphics[scale=0.65]{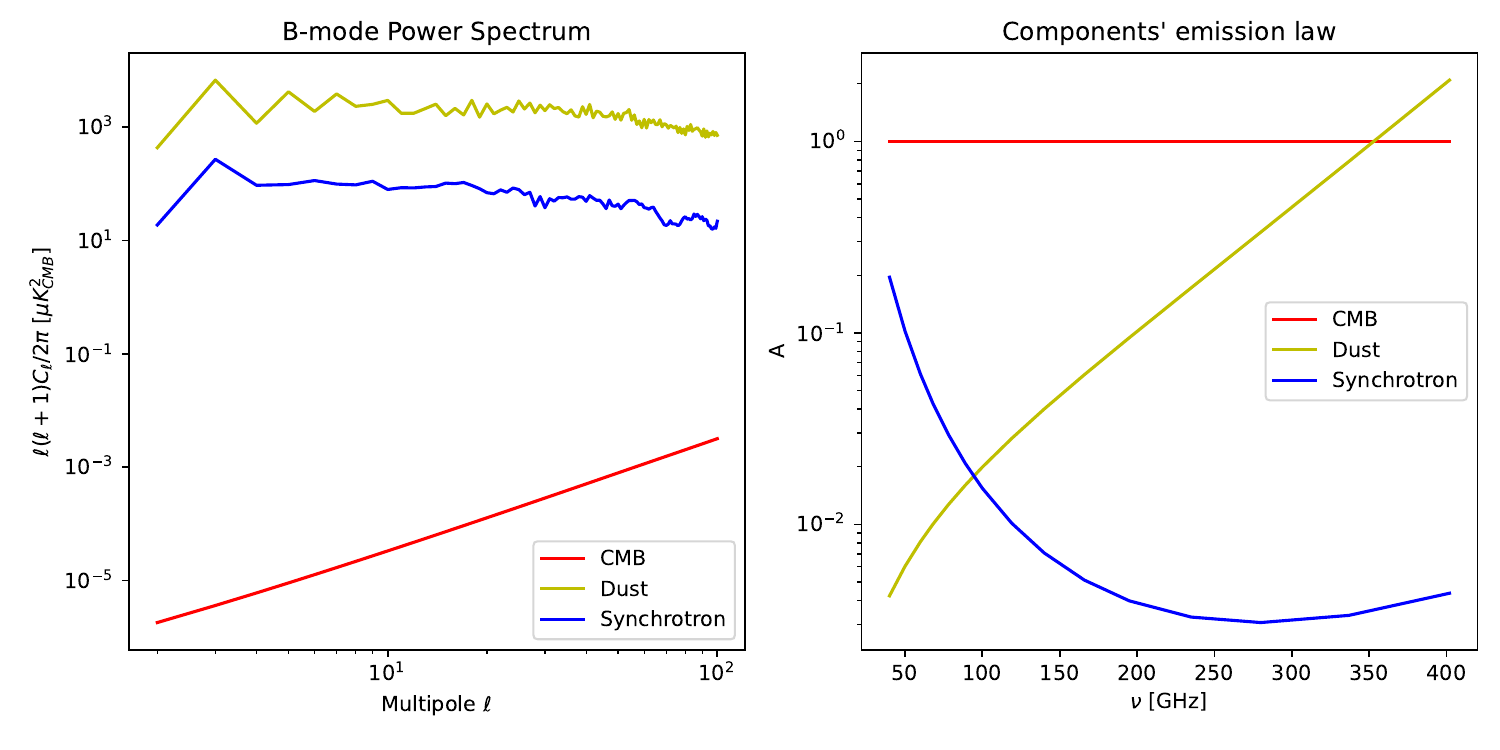}
\caption{\label{fig:data:1} B-mode power spectra computed from template maps at different frequencies: 353~GHz for dust and 23~GHz for synchrotron (left) and the mixing matrix representing the emission laws of the components included in our simulations (right).}
\end{figure}

For the CMB we worked with simulated maps with a tensor-to-scalar ratio $r=0$, hence with no primordial B-modes. We assume gaussian noise with rms specified by the detector sensitivities in table~\ref{tab:lb_rotated}. To describe the dust and synchrotron emission we implement the d0 and s0 models from PySM~\cite{Panexp_2025, Zonca_2021, Thorne_2017}. These correspond to foreground models with no emission spectra variation with respect to the direction on the sky. We work with simulated IQU maps, first converted into B-mode polarization maps and then masked with one of Planck's apodized galactic masks, GAL060 \footnote{\url{http://pla.esac.esa.int/pla/aio/product-action?MAP.MAP_ID=HFI_Mask_GalPlane-apo0_2048_R2.00.fits}}, with an unmasked sky fraction $f_{\text{sky}}\sim 0.6$. In figure~\ref{fig:data:1} we show the power spectra and the mixing matrix representing the emission laws of the three components included in our simulations, normalized to the reference frequencies of the two templates in polarization inside PySM: 353~GHz for dust and 23~GHz for synchrotron.

\FloatBarrier

\section{Power spectrum estimation conditioned by the bispectrum model}
\label{Power_estimation}
\subsection{The multivariate Edgeworth expansion}
\label{sec:MEE}
As mentioned in section~\ref{sec:framework}, the assumptions of the framework implemented within SMICA are
\begin{itemize}
    \item Statistical isotropy of the signal
    \item Gaussianity of the components
    \item Statistical independence between components
    \item Uncorrelated noise in frequency
\end{itemize}
The assumption of Gaussianity of the signal is a good approximation for the CMB~\cite{Planck_2018, Jung_2025} while for the foregrounds it is not true. Neglecting non-Gaussianity in the analysis is therefore a suboptimal choice, even if it does not bias the CMB estimation. In the case of polarization, the main foregrounds are dust and synchrotron radiation~\cite{LB_2025, Dickinson_2016}, whose non-Gaussianity has been studied in the literature, both in terms of $n$-point correlators~\cite{Jung_2018, Coulton_2019} and in terms of their impact on CMB analysis~\cite{Hill_2018, Cabezas_2023, Cabezas_2025, Doohan_2025}.

In order to incorporate this extra information in the SMICA model, which is a component separation method that works in the harmonic space, it is a natural step to relax the assumption of Gaussianity  to include higher order $n$-point correlators. We make use of the Multivariate Edgeworth Expansion (MEE)~\cite{Juszkiewicz_1995, Amendola_1996, Taylor_2001, Bartolo_2012,  Hall_2022}: a generic probability density function (PDF) $P(a_{lm}^d)$, describing the distribution of the spherical harmonic coefficients of the sky in temperature T, E-mode or B-mode polarization independently, at a given frequency indicated by the index $d$, can be expanded around a multivariate gaussian $G(a^d_{lm}| \textbf{C})$ as
\begin{align}
\label{eq:edge:1}
   P(a^d_{lm}) =& \ G(a^d_{lm}|\textbf{C}) \Big( 1 + \frac{1}{6} \langle a^{d_1}_{l_1m_1}  a^{d_2}_{l_2m_2} a^{d_3}_{l_3m_3} \ \rangle h_{I_1I_2I_3}(a^d_{lm} , \textbf{C}) \nonumber\\
   & + \frac{1}{24} \langle a^{d_1}_{l_1m_1}  a^{d_2}_{l_2m_2} a^{d_3}_{l_3m_3} a^{d_4}_{l_4m_4}\rangle \ h_{I_1I_2I_3I_4}(a^d_{lm} , \textbf{C} ) \nonumber\\
   & + \frac{1}{72} \langle a^{d_1}_{l_1m_1}  a^{d_2}_{l_2m_2} a^{d_3}_{l_3m_3} \rangle  \langle a^{d_4}_{l_4m_4}  a^{d_5}_{l_5m_5} a^{d_6}_{l_6m_6} \rangle \ h_{I_1I_2I_3I_4I_5I_6}(a^d_{lm} , \textbf{C} ) + ...\Big),
\end{align}
where
\begin{align}
\label{eq:edge:2}
    h_{I_1I_2...}(a^d_{lm}, \textbf{C}) = (-1)^r G^{-1}(a^{d}_{lm}| \textbf{C}) \frac{\partial^r}{\partial a^{d_1}_{l_1m_1} \partial a^{d_2}_{l_2m_2}...}  G(a^{d}_{lm}| \textbf{C})
\end{align}
are the Hermite tensors  of order $r$ (the number of subscripts), the multivariate generalizations of the Hermite polynomials. The MEE gives a good approximation for a generic distribution only in the case of weak non-Gaussianity, since the assumption behind it is that higher-order correlators are subdominant with respect to the ones included in the expansion. Indeed, even if it is well normalized, the expansion can produce negative non-physical values if the variance and skewness  are too large. In this work we will stick to the first-order term including the 3-point correlator. In appendix~\ref{app:MEE} we show that by computing the derivatives and using the definitions given in section~\ref{sec:nonG}, this expansion is equivalent to
\begin{align}
\label{eq:edge:3}
    P(a^d_{lm}|\textbf{C},\textbf{B}) = G(a^d_{lm}|\textbf{C}) \Big[ 1 + \langle \textbf{B}, \widehat{\textbf{B}} - \textbf{B}^{\mathrm{lin \ corr}}\rangle \Big],
\end{align}
where $\langle \textbf{B}^I, \textbf{B}^J \rangle $ is a mathematically well-defined inner product between two bispectra  defined by~\cite{Bucher_2016} 
\begin{align}
	\label{eq:edge:4}
	\langle \textbf{B}^I, \textbf{B}^J \rangle =& \sum_{l_1\leq l_2 \leq l_3} \sum_{d_i,d'_i} (-1)^{l_1+l_2+l_3}(B^I)^{d_1d_2d_3}_{l_1l_2l_3} \left(\mathrm{Var}^{-1}(\textbf{B})\right)^{d_1d_2d_3,d'_1d'_2d'_3}_{l_1l_2l_3} (B^J)^{d'_1d'_2d'_3}_{l_1l_2l_3} ,
\end{align}
and $\mathrm{Var}(\textbf{B})$ is the bispectrum variance, which in the weakly non-gaussian case is equal to
\begin{align}
	\label{eq:edge:5}
	\left(\mathrm{Var}(\textbf{B})\right)^{d_1d_2d_3, d'_1d'_2d'_3}_{l_1l_2l_3} = g_{l_1l_2l_3} C^{d_1d'_1}_{l_1} C^{d_2d'_2}_{l_2} C^{d_3d'_3}_{l_3}.
\end{align}
The factor $g_{l_1l_2l_3}$ is respectively 6 if all $l$'s are equal, 2 if two of them are equal and 1 if all of them are different.

In eq.~\eqref{eq:edge:3} $\textbf{B}$ is the model bispectrum defined by 
\begin{align}
	\label{eq:edge:6}
	B^{d_1d_2d_3}_{l_1l_2l_3} =& \ \langle a^{d_1}_{l_1m_1}a^{d_2}_{l_2m_2} a^{d_3}_{l_3m_3}\rangle \nonumber\\
        =&\left\langle \Big( \sum_{c_1=1}^{N_{\mathrm{comp}}} A^{d_1 c_1}s^{c_1}_{l_1m_1} + n^{d_1}_{l_1m_1}\Big)\Big( \sum_{c_2=1}^{N_{\mathrm{comp}}} A^{d_2 c_2}s^{c_2}_{l_2m_2} + n^{d_2}_{l_2m_2}\Big)\Big( \sum_{c_3=1}^{N_{\mathrm{comp}}} A^{d_3 c_3}s^{c_3}_{l_3m_3} + n^{d_3}_{l_3m_3}\Big)\right\rangle\nonumber\\
        =&\sum_c A^{d_1 c}A^{d_2 c}A^{d_3 c}\langle s^{c}_{l_1m_1}s^{c}_{l_2m_2}s^{c_3}_{l_3m_3}\rangle + \langle n^{d_1}_{l_1m_1}n^{d_2}_{l_2m_2}n^{d_3}_{l_3m_3}\rangle\nonumber\\
        =& \sum_{m_i}\begin{pmatrix}
	    l_1 & l_2 & l_3 \\ m_1 & m_2 & m_3
	\end{pmatrix} \sum_c A^{d_1c} A^{d_2c} A^{d_3c} B^{c}_{l_1l_2l_3},
\end{align}
where $A^{dc}$ is the $d$-th element of the mixing matrix of the $c$-th sky component with a non-zero $3$-point correlator $B^c_{l_1l_2l_3}\propto\langle s^c_{l_1m_1}s^c_{l_2m_2}s^c_{l_3m_3}\rangle$. Here we have used all of our assumptions: gaussian and uncorrelated noise, uncorrelated components, statistical isotropy, and weak non-Gaussianity for the foreground components.

While $\widehat{\textbf{B}}$ is the observed bispectrum defined in eq.~\eqref{eq:nonG:7}, $\textbf{B}^{\mathrm{lin \ corr}}$ is the so called linear correction term which is defined by
\begin{align}
\label{eq:edge:7}
	\Big(B^{\mathrm{lin \ corr}}\Big)^{d_1d_2d_3}_{l_1l_2l_3} = \sum_{m_i}\begin{pmatrix}
	    l_1 & l_2 & l_3 \\ m_1 & m_2 & m_3
	\end{pmatrix}\Big[\langle a^{d_1}_{l_1m_1}a^{d_2}_{l_2m_2} \rangle a^{d_3}_{l_3m_3} + \text{2 permutations}\Big].
\end{align}
This term is needed in the case of anisotropic noise or incomplete sky coverage in order to reduce the variance~\cite{Planck_2013}. Since we work under the assumption of isotropy, this term is assumed to be a small correction to the bispectrum variance~\cite{Bucher_2016}, hence we use the distribution
\begin{align}
\label{eq:edge:8a}
    P(a^d_{lm}|\textbf{C},\textbf{B}) = G(a^d_{lm}|\textbf{C}) \Big[ 1 + \langle \textbf{B}, \widehat{\textbf{B}}\rangle \Big].
\end{align}
In order to be a well-defined probability, it needs to be normalized:
\begin{align}
\label{eq:edge:8b}
    \int  d\textbf{a} \ P(a^d_{lm}) =& \int d \textbf{a} \ G(a^d_{lm}|\textbf{C}) \Big[ 1 + \langle \textbf{B}, \widehat{\textbf{B}}\rangle \Big]\nonumber\\
    =& \int d \textbf{a} \  G(a^d_{lm}|\textbf{C}) + \int d \textbf{a} \ G(a^d_{lm}|\textbf{C})\langle \textbf{B}, \widehat{\textbf{B}}\rangle \nonumber\\
    =& 1 + \int d \textbf{a} \ G(a^d_{lm}|\textbf{C})\langle \textbf{B}, \widehat{\textbf{B}}\rangle\nonumber\\
    =& 1 + \int d \textbf{a} \ G(a^d_{lm}|\textbf{C}) \sum_{l_i,d_i,d'_i} B^{d_1d_2d_3}_{l_1l_2l_3}\left(\mathrm{Var}^{-1}(\textbf{B})\right)_{l_1l_2l_3}^{d_1d_2d_3,d'_1d'_2d'_3}a^{d_1}_{l_1m_1}a^{d_2}_{l_2m_2}a^{d_3}_{l_3m_3} \nonumber\\
    =& 1,
\end{align}
where the last integral vanishes because the integrand is an odd function of $a^d_{lm}$. In eq.~\eqref{eq:edge:8b} we have defined $d\textbf{a}=\prod_{d=1}^{N_{\text{freq}}}\prod_{l=0}^{+\infty}\prod_{m=-l}^l da^d_{lm}$.
The probability in eq.~\eqref{eq:edge:8a} also needs to be positive 
\begin{align}
\label{eq:edge:9}
    P(a^d_{lm}|\textbf{C}, \textbf{B}) = G(a^d_{lm}|\textbf{C}) \Big[ 1 + \langle \textbf{B}, \widehat{\textbf{B}}\rangle \Big]&\geq 0, \nonumber\\
     \implies|\langle \textbf{B}, \widehat{\textbf{B}}\rangle | = |\textbf{B}| |\widehat{\textbf{B}}| |\cos(\theta)| &\leq 1,
\end{align}
where $|\textbf{B}|$ is the norm in the bispectrum space defined by the inner product in eq.~\eqref{eq:edge:4}, which also defines $\theta$ as the "angle" between the two bispectra. This will be satisfied for every angle $\theta$ if $|\textbf{B}| |\widehat{\textbf{B}}|\leq1$, which is consistent with the first-order expansion of the MEE as the bispectrum should be a small correction to the gaussian distribution in the limit
\begin{align}
\label{eq:edge:10}
    |\textbf{B}| |\widehat{\textbf{B}}| \ll  1.
\end{align}
This effectively corresponds to the weakly non-Gaussian limit, i.e., the squared bispectrum being small with respect to the Gaussian bispectrum variance.

Now, given one observed sky at different frequencies $\widehat{a}^d_{lm}$, we can write a new likelihood
\begin{align}
\label{eq:edge:11}
    \mathcal{L}(\textbf{C}, \textbf{B}| \widehat{a}^d_{lm}) = G(\widehat{a}^d_{lm}|\textbf{C}) \Big[ 1 + \langle \textbf{B}, \widehat{\textbf{B}}\rangle \Big]
\end{align}
and consequently the negative log-likelihood
\begin{align}
\label{eq:edge:12}
    -\log\Big(\mathcal{L}(\textbf{C}, \textbf{B}| \widehat{\textbf{C}}, \widehat{\textbf{B}}) \Big)=&-\log\Big(\mathcal{L}_G(\textbf{C}| \widehat{\textbf{C}}) \Big) -\log\Big(\mathcal{L}_{NG}(\textbf{C}, \textbf{B}|\widehat{\textbf{B}}) \Big)\nonumber\\
    =&\frac{1}{2}\sum_l (2l+1)D(\widehat{\textbf{C}}_l,\textbf{C}(\vec{\theta})_l) - \log\Big( 1 + \langle \textbf{B}, \widehat{\textbf{B}}\rangle \Big) +\text{const}.
\end{align}
which corresponds to effectively enlarging the parameter space to
\begin{align}
\label{eq:edge:13}
    \vec{\theta}= \{ A^{dc}, C^c_l, \sigma^d_l, B^c_{l_1l_2l_3}\}.
\end{align}
In eq.~\eqref{eq:edge:12} we have defined the original SMICA likelihood as $\mathcal{L}_G$ and the corrective non-gaussian term as $\mathcal{L}_{NG}$. However, this likelihood is not suitable to fit for the bispectrum as well. The MEE is a first-order expansion around a gaussian distribution, hence it is only linear in the bispectrum: the non-gaussian term in eq.~\eqref{eq:edge:13} can be rewritten as
\begin{align}
\label{eq:edge:13b}
1+|\textbf{B}||\widehat{\textbf{B}}|\cos(\theta).
\end{align}
Therefore, if $\cos(\theta)>0$ ($<0$), the probability is higher for higher (lower) values of $|\textbf{B}|$, pushing it to the limit of the domain defined by $|\textbf{B}||\widehat{\textbf{B}}|\leq 1$, which is not consistent with the weakly non-gaussian assumptions we made about the bispectrum. In order to deal with this limitation, we have chosen to use a bispectrum model for the foreground components which we define in the following section.

\subsection{Bispectrum model}
\label{subsec:datamodel}
\begin{figure}[tbp]
\hspace*{-2cm} 
\centering 
\includegraphics[scale=0.65]{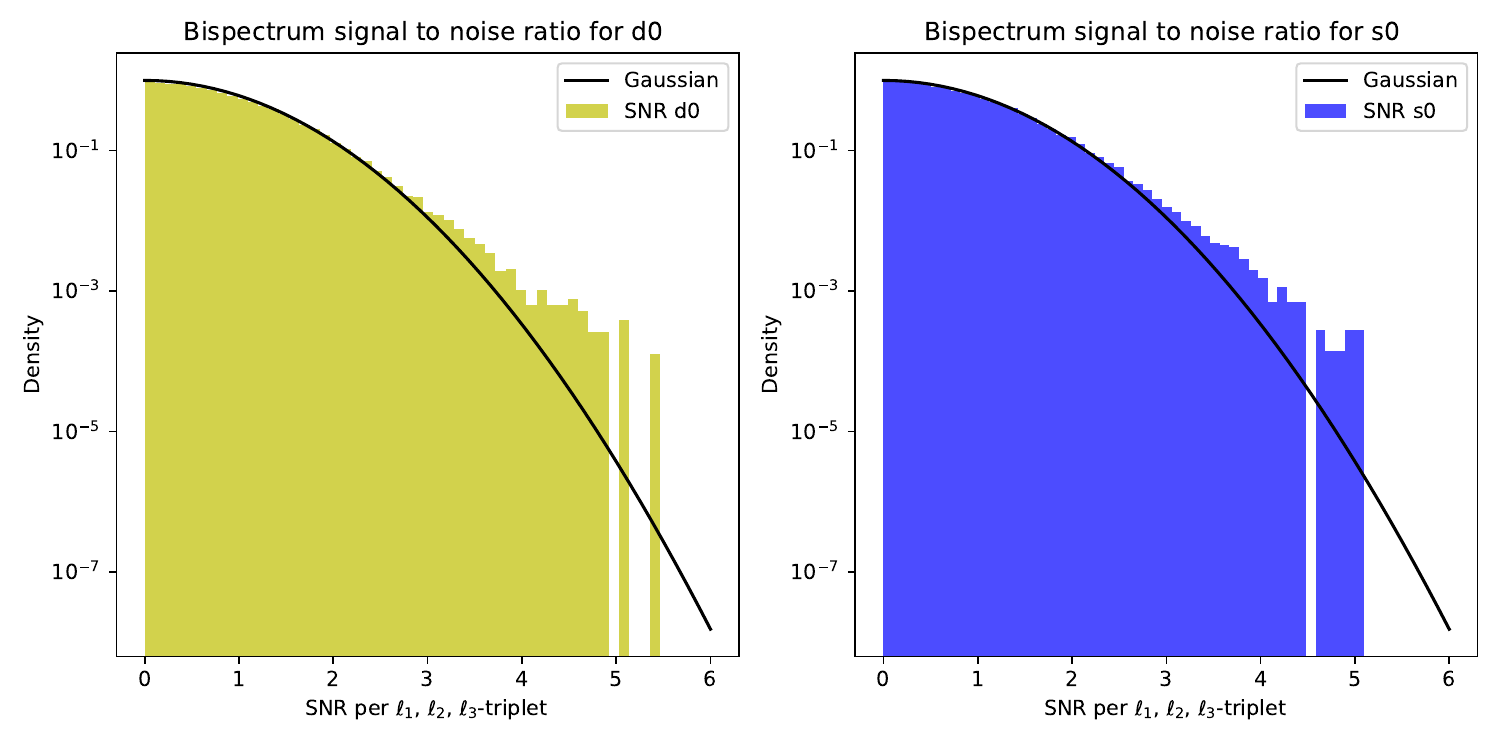}
\caption{\label{fig:datamodel:2} The signal-to-noise ratio of the bispectrum for the two masked foreground maps d0 and s0.}
\end{figure}

SMICA uses a preconditioned conjugate gradient descent algorithm that fits for $\mathcal{O}(300)$ parameters. For our dataset with $\ell_{\text{max}} = 100$ the bispectrum of one component is instead an object of $(\ell_{\text{max}}-2)(\ell_{\text{max}}-1)(\ell_{\text{max}})/6=166650$ elements. Not only is it a huge numerical difficulty to fit for $\mathcal{O}(10^{5})$ parameters, but we are also treating the bispectrum as sub-dominant and therefore noisy. Moreover, as already said in section~\ref{sec:MEE}, the bispectrum appears only linearly in the MEE as it is a first-order approximation, thus, we decided to find a physically motivated bispectrum model for the two foreground components. As explained in section~\ref{sec:nonG}, the observed bispectrum of one realization is greatly affected by the variance, which means that in order to model the true $3$-point correlator of dust and synchrotron we need to look at the signal to noise ratio defined as 
\begin{align}
    \label{eq:datamodel:1}
    \mathrm{SNR} =& \sqrt{\frac{B^2_{l_1l_2l_3}}{g_{l_1l_2l_3}C_{l_1}C_{l_2}C_{l_3}}}.
\end{align}
In figure \ref{fig:datamodel:2} we show the histogram of this quantity for the two masked foreground templates used in our simulations compared with what one expects to see for a perfectly gaussian map. It is clear that we are in the limit of weak non-Gaussianity due to the very low signal-to-noise ratio characterizing the bispectrum of the two masked foregrounds. 

\begin{figure}[tbp]
\centering 
\includegraphics[scale=0.55]{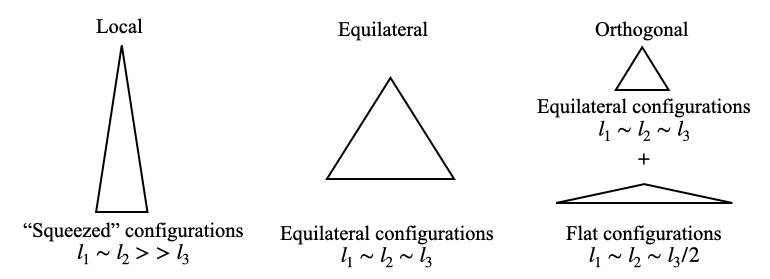}
\caption{\label{fig:datamodel:3} The three bispectrum models implemented to describe the bispectra of d0 and s0. These three models peak in the configurations indicated.}
\end{figure}

It is important to note that SMICA assumes that the foregrounds are realizations of stochastic processes, hence that the bispectrum $B_{l_1l_2l_3}$ defined by $\langle a_{l_1m_1}a_{l_2m_2}a_{l_3m_3}\rangle$ is a well defined quantity that differs from the observed bispectrum of a single realization, which is dominated (in the limit of weak non-Gaussianity) by the gaussian variance $g_{l_1l_2l_3}C_{l_1}C_{l_2}C_{l_3}$. This means that what we are trying to model is the small signal on top of this gaussian distribution.  
Inspired by primordial non-Gaussianity studies, in order to model the "true" $3$-point correlator, we decided to build a bispectrum based on 3 fundamental templates constructed via the power spectrum of the two masked foregrounds:
\begin{itemize}
    \item The local bispectrum~\cite{Gangui_1994}, which peaks in the "squeezed" configurations $\ell_1\ll \ell_2\sim \ell_3$, modeled as 
    \begin{align}
    \label{eq:datamodel:2}
        B^{\text{loc}}_{l_1l_2l_3} = -C_{l_1}C_{l_2} + \text{2 perm.}.
    \end{align}
    
    \item The equilateral bispectrum~\cite{Creminelli_2006}, which peaks in the "equilateral" configurations $\ell_1\sim\ell_2\sim \ell_3$, modeled as
    \begin{align}
\label{eq:datamodel:3}
        B^{\text{eq}}_{l_1l_2l_3} = \Big[C_{l_1}C_{l_2} + \text{2 perm.}\Big] + 2 C_{l_1}^{2/3}C_{l_2}^{2/3}C_{l_2}^{2/3} - \Big[C_{l_1}C_{l_2}^{2/3}C_{l_3}^{1/3} + \text{5 perm.}\Big].
    \end{align}
    \item The orthogonal bispectrum~\cite{Senatore_2010}, constructed to be orthogonal to the equilateral bispectrum, modeled as
    \begin{align}
    \label{eq:datamodel:4}
        B^{\text{ort}}_{l_1l_2l_3} = 3\Big[C_{l_1}C_{l_2} + \text{2 perm.}\Big] + 8 C_{l_1}^{2/3}C_{l_2}^{2/3}C_{l_2}^{2/3} -3 \Big[C_{l_1}C_{l_2}^{2/3}C_{l_3}^{1/3} + \text{5 perm.}\Big].
    \end{align}
\end{itemize}
In these definitions, $C_l$ represents the power spectrum of each masked foreground, respectively. The masks used are Planck’s apodized galactic masks GAL020-GAL099 where 20-99 is the fraction of the sky that has been left unmasked. The spectra are pseudo-spectra corrected by a $\frac{1}{f_{\mathrm{sky}}}$ factor. The models we have defined have a signal-to-noise ratio with respect to the gaussian variance that peaks in the configurations we show in figure~\ref{fig:datamodel:3}. Via these three templates, we have decomposed the bispectrum of dust and synchrotron as 
\begin{align}
    \label{eq:datamodel:5}
    B^{\text{foreground}}_{l_1l_2l_3} = \sum_I f^I_{\text{NL}} B^I_{l_1l_2l_3},
\end{align}
where
\begin{align}
    \label{eq:datamodel:6}
    f^I_{\text{NL}} =& \sum_J (F^{-1})^{IJ}\langle \textbf{B}^J, \textbf{B}^{obs}\rangle, \nonumber\\
    F^{IJ} =& \langle \textbf{B}^I, \textbf{B}^J\rangle.
\end{align}
This effectively means decomposing the observed bispectrum of d0 and s0 into a basis in the bispectrum space. Although the 3 models we have defined are not a complete basis in this space, they cover a huge variety of physical processes. Our approach is the best we can do given the lack of theoretical work regarding the bispectra of the foregrounds. One way to improve this model would be to use full-sky simulations of dust and synchrotron capable of producing different realizations of non-gaussian maps based on the same physics, which is a constantly developing topic in the cosmology community~\cite{Thorne_2021, Parente_2025, Waelkens_2009, Spinelli_2018, Diao_2025}, to compute the average bispectrum over a distribution of maps, but it is beyond the scope of this work.

\begin{figure}[tbp]
\centering 
\includegraphics[scale=0.7]{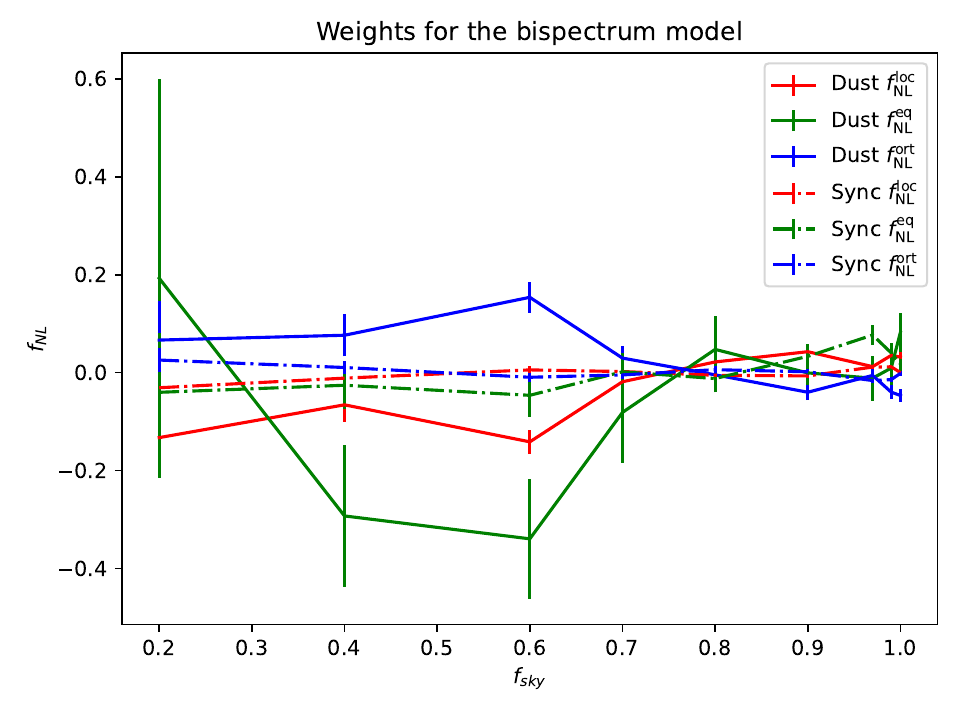}
\caption{\label{fig:datamodel:4} The $f_{\text{NL}}$ values for the three shapes as a function of $f_{\text{sky}}$, both for dust and synchrotron, with $1\sigma$ Fisher error bars.}
\end{figure}

We noticed that the results obtained were highly mask-dependent, as we show in figure~\ref{fig:datamodel:4}. In order to assess whether the scatter was caused by the absence of the linear correction term, we computed it for all the masks implemented, both for dust and synchrotron. Unfortunately, as we have access to only one realization of the foregrounds, the maps used to compute the linear correction term are gaussian statistically isotropic maps built via the power spectrum of the foregrounds. This implies that this linear correction can only reduce the mask effect and not the intrinsic anisotropy of the foregrounds, a similar problem found in other works~\cite{Coulton_2019}. With this term, we obtained the same behaviour as shown in figure~\ref{fig:datamodel:4}, with a noticeable effect only for $f_{\text{sky}}=0.20$, as there the mask covers most of the sky and becomes the dominant effect. The weights are more stable for higher $f_{\text{sky}}$, while, for our mask with $f_{\text{sky}}=0.60$, there is a clear jump in values, as it is the mask that cuts off most of the signal of the foregrounds. In order to model the true bispectrum, we decided to use the weights obtained at $f_{\text{sky}}=0.90$ (without the very bright galactic center), rescaled by a factor $f^{3/2}$ which is computed as 
\begin{align}
    \label{fsky}
    f = \overline{\left( \frac{C_l(f_{\text{sky}}=0.6)}{C_l(f_{\text{sky}}=0.9)} \right)}.
\end{align}
Rescaling by $f_{\text{sky}}^{3/2}$ would have been a good approximation if the signal was not as mask-dependent, but here it is better to use $f^{3/2}$. 
The two factors obtained are $f=0.18$ for dust and  $f=0.33$ for synchrotron. What we have obtained for the local, equilateral and orthogonal templates is
\begin{align}
    \label{eq:datamodel:7}
    &f_{NL}(d0, f_{\text{sky}}=0.9) = \{ 0.043, -5.1\cdot 10^{-4},  -0.040\}, \nonumber\\
    &f_{NL}(s0, f_{\text{sky}}=0.9) = \{  -0.0067, 0.033, 0.0018\}.
\end{align}
The significance of these numbers has to be compared with the expected fisher error, which is defined by
\begin{align}
    \label{eq:datamodel:7b}
    \Delta f^I_{\text{NL}} = \sqrt{\left(F^{-1}\right)^{II}},
\end{align}
which is
\begin{align} 
    \label{eq:datamodel:7c} 
    &\Delta f_{NL}(d0, f_{\text{sky}}=0.9) = \{ 0.012, 0.055, 0.015\}, \nonumber\\
    &\Delta f_{NL}(s0, f_{\text{sky}}=0.9) = \{  0.0044, 0.025, 0.0056\}.
\end{align}
For the two foregrounds, the ratio between $f_{\text{NL}}$ and its minimum variance error is 
\begin{align}
    \label{eq:datamodel:7d} 
    &f_{NL}/\Delta f_{NL}(d0, f_{\text{sky}}=0.9) = \{ 3.7, 0.0091, 2.7\}, \nonumber\\
    &f_{NL}/\Delta f_{NL}(s0, f_{\text{sky}}=0.9) = \{  1.5, 1.3, 0.31\}.
\end{align}
We see that the values indicate a detection of the models analyzed, with the most significant value for both foregrounds for the local shape, coherent with the previous results~\cite{Jung_2018, Coulton_2019}.

The total model bispectrum is therefore
\begin{align}
    \label{eq:datamodel:8}
    B_{l_1l_2l_3}^{d_1d_2d_3} =& A^{d_1,\text{dust}}A^{d_2,\text{dust}}A^{d_3,\text{dust}} f^{3/2}(d0) B^{\text{dust}}_{l_1l_2l_3}(f_{\text{sky}}=0.9) \nonumber\\
    &+A^{d_1,\text{sync}}A^{d_2,\text{sync}}A^{d_3,\text{sync}} f^{3/2}(s0)B^{\text{sync}}_{l_1l_2l_3}(f_{\text{sky}}=0.9).
\end{align}
With this model, we are approximating the true non-gaussian signal via a decomposition into three standard and fairly general shapes. The weights of this decomposition have been estimated from the bispectrum of the input map. In order to estimate the benefit of our approach, in the case of real observations it is possible to estimate these parameters directly from the data, as we will discuss later in section~\ref{sec:mdmcbb}. 

As we have access to only one realization of the foreground maps, it is impossible to obtain an unbiased estimate of the goodness of fit of our model. We highlight that what we model here is the $3$-point correlator of the foregrounds averaged over infinitely many realizations, inferred from a single observed sky — an estimate that is strongly affected by the Gaussian variance in the weak non-Gaussianity regime. Moreover, the intrinsic statistical anisotropy of the foregrounds makes our analysis strongly mask-dependent, and limits our ability to validate the model against simulations, as such anisotropy cannot be properly emulated.

\subsection{Parameters estimation}
\label{sec:res}
We performed a power spectrum and spectral parameters estimation that uses both the gaussian and first-order non-gaussian information in the analysis. In order to do this, we have modified our algorithm for our new log-likelihood
\begin{align}
\label{eq:res:1}
    -\log(\mathcal{L}(\textbf{C}|\widehat{\textbf{C}},\textbf{B},\widehat{\textbf{B}}) )=& \frac{1}{2}\sum_l (2l+1)D(\widehat{\textbf{C}}_l,\textbf{C}(\vec{\theta})_l) + \log\Big( 1 + \langle \textbf{B}, \widehat{\textbf{B}}\rangle \Big) +\text{const.}\nonumber\\
    =& \frac{1}{2}\sum_l (2l+1)D(\widehat{\textbf{C}}_l,\textbf{C}(\vec{\theta})_l) \nonumber\\
    &+ \log\Big( 1 + \sum_{l_i,d_i,d'_i} \sum_c A^{c,d_1}A^{c,d_2}A^{c,d_3}B^c_{l_1l_2l_3} \left(\mathrm{Var}^{-1}(\textbf{B})\right)^{d_1d_2d_3,d'_1d'_2d'_3}_{l_1l_2l_3}\widehat{B}_{l_1l_2l_3}^{d'_1d'_2d'_3}\rangle \Big) +\text{const.},
\end{align}
where $B^c_{l_1l_2l_3}$ are the two bispectrum models defined in section~\ref{subsec:datamodel} and
\begin{align}
\label{eq:res:w}
    &\left(\mathrm{Var}^{-1}(\textbf{B})\right)^{d_1d_2d_3,d'_1d'_2d'_3}_{l_1l_2l_3} = \frac{1}{g_{l_1l_2l_3}}(C^{-1}_{l_1})^{d_1d'_1}(C^{-1}_{l_2})^{d_2d'_2}(C^{-1}_{l_3})^{d_3d'_3},\nonumber\\
    &C_{l}^{dd'}(\vec{\theta}) = A^{dc}C_{l}^cA^{d',c}+\delta^{dd'}(\sigma_l^d)^2.
\end{align}
We have then run it on the set of 400 simulations described in section~\ref{sec:sim} to fit for the set of parameters $\vec{\theta} = \{ A^{dc}, C^c_l, \sigma^d_l\}$ and compared the results with the one obtained by running with the original likelihood. Unfortunately, the results were identical up to a relative factor of $\mathcal{O}(10^{-7})$, well within the error bars of the algorithm itself. 

\begin{figure}[tbp]
\hspace*{-2cm} 
\centering 
\includegraphics[scale=0.65]{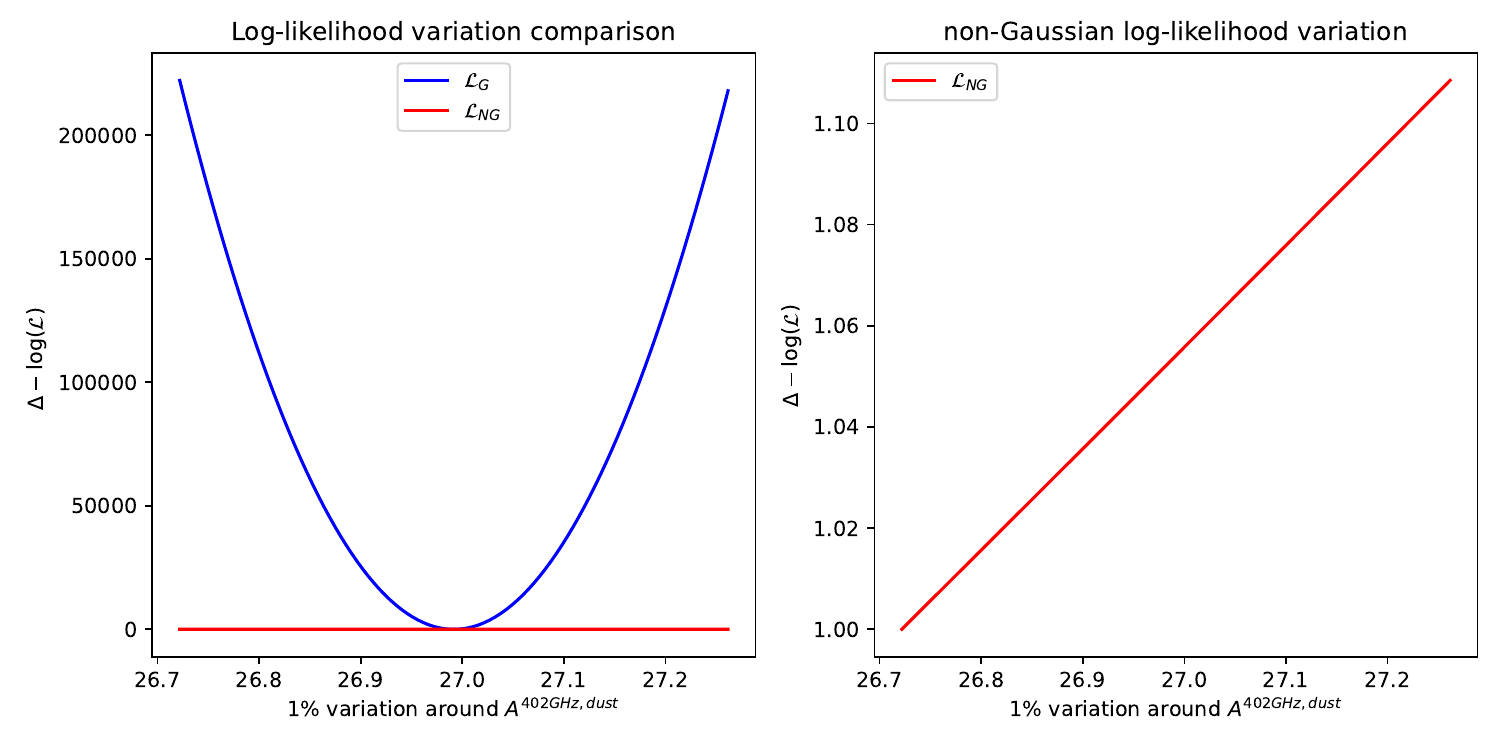}
\caption{\label{fig:res:2} Comparison between the original gaussian likelihood and the non-Guassian corrective term for the mixing matrix component $A^{402\text{GHz, dust}}$.}
\end{figure}

The reason behind this result is the scale over which the non-gaussian term we added to the log-likelihood varies. In figure~\ref{fig:res:2} we have plotted how the two terms defined in eq~\eqref{eq:edge:12} vary around the minimum found by gaussian SMICA for the last element of the dust mixing matrix which corresponds to 402~GHz. Although, of course, for a likelihood describing the distribution of $\mathcal{O}(10^2)$ parameters, the variation for a single one of them is not a complete description of the full shape, it gives a clear idea of the amount of information that it is adding to the gaussian likelihood. We performed the same check for different parameters and obtained the same result: the non-gaussian contribution varies on much larger scales than the gaussian leading-order term, and therefore does not improve the constraining power of the parameters.

The physical explanation traces back to the assumptions we have made in this work. One cause could be our limited knowledge about the foregrounds' bispectrum model, since no well-motivated theoretical model exists. Moreover, truncating the MEE to the first order is a strong assumption on the non-Gaussianity of the foregrounds: it is possible that higher-order correlators actually dominate over the bispectrum. We would like to point out that we also tried to reduce the number of available frequencies, specifically using 5 frequency channels, in order to test our approach in a less constrained setting, but we obtained similar results.

Adding the bispectrum information does not improve the estimation of the power spectrum; therefore, one could think of first estimating the covariance matrix $\textbf{C}$ via the standard version of SMICA and then using the MEE to estimate the bispectrum. Following this direction shows an important limit of the MEE: as already discussed in section~\ref{sec:MEE}, it is a first-order approximation of a non-gaussian distribution and the bispectrum only appears linearly. Therefore, if we minimize the non-gaussian negative log-likelihood
\begin{align}
    \label{eq:ind:1}
    -\log\Big(\mathcal{L}_{NG}(\textbf{B}|\textbf{C},\widehat{\textbf{B}})\Big) =& -\log(1+\langle \textbf{B},\widehat{\textbf{B}}\rangle)\nonumber\\
    =& -\log(1+\sum_{l_1\leq l_2 \leq l_3}\sum_{d_i,d'_i} B^{d_1d_2d_3}_{l_1l_2l_3} \left(Var(\textbf{B})\right)^{d_1d_2d_3,d'_1d'_2d'_3}_{l_1l_2l_3} \widehat{B}^{d'_1d'_2d'_3}_{l_1l_2l_3} ),
\end{align}
we would get a gradient for the bispectrum pointing in one constant direction. This is still mathematically correct, as the bispectrum is defined only in the domain $|\textbf{B}||\widehat{\textbf{B}}|\leq 1$, see eq.~\eqref{eq:edge:9}, but it is not consistent with the assumption of weak non-Gaussianity and it is dependent on the 
starting point of the gradient descent.

\FloatBarrier

\section{Independent power spectrum and conditioned bispectrum estimation}
\label{sec:mdmcbb}
\subsection{The likelihood}
\label{sec:mdmcbb:lik}
The results of section~\ref{sec:res} have an important consequence: an independent estimation of the power spectrum and conditionally the bispectrum is well motivated. Of course, this is an approximation since the two correlators are not really independent. If we compute their covariance for a perfectly gaussian distribution, we get
\begin{align}
    \label{eq:like:1}
    \langle \textbf{C} \ \textbf{B} \rangle \sim \langle a^5\rangle =0,
\end{align}
but this only means that they are not correlated: they are both observables constructed with the same random variable $a_{lm}$ and are therefore not independent. 

Given what we said at the end of the previous section, after estimating the power spectra with the original SMICA method, we decided to adopt a new likelihood to describe the bispectrum. Indeed, the bispectrum distribution alone is well approximated by a gaussian distribution ~\cite{Sohn_2023}
\begin{align}
    \label{eq:like:2}
    \mathcal{L}(\textbf{B}|\widehat{\textbf{B}}) =& \ G\big(\widehat{\textbf{B}}|\mathrm{Var}(\textbf{B}) = g_{l_1l_2l_3}C_{l_1}C_{l_2}C_{l_3},\overline{\textbf{B}}=\textbf{B}\big)\nonumber\\
    =&\prod_{l_1\leq l_2\leq l_3} \frac{1}{\sqrt{2\pi}\sqrt{g_{l_1l_2l_3}C_{l_1}C_{l_2}C_{l_3}}}\exp \Big[ -\frac{1}{2} \big(\widehat{B}_{l_1l_2l_3}-B_{l_1l_2l_3}\big) \frac{1}{g_{l_1l_2l_3}C_{l_1}C_{l_2}C_{l_3}}\big(\widehat{B}_{l_1l_2l_3}-B_{l_1l_2l_3}\big)\Big] \nonumber\\
    =& \prod_{l_1\leq l_2\leq l_3} \frac{1}{\sqrt{2\pi}\sqrt{g_{l_1l_2l_3}C_{l_1}C_{l_2}C_{l_3}}}\exp \Big( -\frac{1}{2}\langle \widehat{\textbf{B}}-\textbf{B},\widehat{\textbf{B}}-\textbf{B}\rangle\Big).
\end{align}
We did not use this likelihood before to include the $3$-point correlator in SMICA since, as already mentioned, the power spectrum and the bispectrum are not independent, and a joint likelihood of the form
\begin{align}
    \label{eq:like:3}
    \mathcal{L}(\textbf{C},\textbf{B}|\widehat{\textbf{C}},\widehat{\textbf{B}})=\mathcal{L}(\textbf{C}|\widehat{\textbf{C}})\mathcal{L}(\textbf{B}|\widehat{\textbf{B}}) = G(\widehat{a}_{lm}| \textbf{C}) G\left(\widehat{\textbf{B}}|\mathrm{Var}(\textbf{B}) = g_{l_1l_2l_3}C_{l_1}C_{l_2}C_{l_3},\overline{\textbf{B}}=\textbf{B}\right)
\end{align}
would have been biased because the correlation between the variables would have been neglected: if we look at the gaussian limit for $\textbf{B}\to 0$, we observe
\begin{align}
    \label{eq:like:4}
     \lim_{B\to0} \mathcal{L}(\textbf{C},\textbf{B}|\widehat{\textbf{C}},\widehat{\textbf{B}}) = G(\widehat{a}_{lm}| \textbf{C}) \prod_{l_1\leq l_2\leq l_3} \frac{1}{\sqrt{2\pi}\sqrt{g_{l_1l_2l_3}C_{l_1}C_{l_2}C_{l_3}}}\exp \Big( -\frac{1}{2}\langle \widehat{\textbf{B}},\widehat{\textbf{B}}\rangle\Big)\neq G(\widehat{a}_{lm}| \textbf{C}).
\end{align}
If we correct the joint likelihood in eq.~\eqref{eq:like:3} for the remaining factor
\begin{align}
    \label{eq:like:5}
     \frac{\lim_{\textbf{B}\to0} \mathcal{L}(\textbf{C},\textbf{B}|\widehat{\textbf{C}},\widehat{\textbf{B}})}{G(\widehat{a}_{lm}| \textbf{C})} =  \prod_{l_1\leq l_2\leq l_3} \frac{1}{\sqrt{2\pi}\sqrt{g_{l_1l_2l_3}C_{l_1}C_{l_2}C_{l_3}}}\exp \Big( -\frac{1}{2}\langle \widehat{\textbf{B}},\widehat{\textbf{B}}\rangle\Big),
\end{align}
we get 
\begin{align}
    \label{eq:like:6}
    \mathcal{L}(\textbf{C},\textbf{B}|\widehat{\textbf{C}},\widehat{\textbf{B}})=G(\widehat{a}_{lm}| \textbf{C}) \exp\Big( -\frac{1}{2}\langle \textbf{B},\textbf{B} \rangle+\langle \textbf{B}, \widehat{\textbf{B}} \rangle\Big),
\end{align}
which in the limit $\textbf{B}\to0$, at first order, is exactly the MEE we obtained before
\begin{align}
    \label{eq:like:7}
    \lim_{\textbf{B}\to 0}\mathcal{L}(\textbf{C},\textbf{B}|\widehat{\textbf{C}},\widehat{\textbf{B}})=G(\widehat{a}_{lm}| \textbf{C}) \Big( 1+\langle \textbf{B},\widehat{\textbf{B}}\rangle +\mathcal{O}(\textbf{B}^2)\Big).
\end{align}
Given all these considerations, we will consider an independent bispectrum likelihood, conditioned by the estimation of the power spectrum by SMICA, generalizing the one in eq.~\eqref{eq:like:2} to the multi-frequency space
\begin{align}
    \label{eq:like:8}
    \mathcal{L}(B|\widehat{B}) &\propto \exp \Big[ -\frac{1}{2}\sum_{l_i,d_i,d'_i} \big(\widehat{B}-B\big)^{d_1d_2d_3}_{l_1l_2l_3} \left(\mathrm{Var}^{-1}(\textbf{B})\right)^{d_1d_2d_3,d'_1d'_2d'_3}_{l_1l_2l_3}\big(\widehat{B}-B\big)^{d'_1d'_2d'_3}_{l_1l_2l_3}\Big] \nonumber\\
    =&\exp \Big( -\frac{1}{2}\langle \widehat{\textbf{B}}-\textbf{B},\widehat{\textbf{B}}-\textbf{B}\rangle\Big)
\end{align}
with the same multi-component bispectrum model of eq.~\eqref{eq:edge:6}
\begin{align}
	\label{eq:like:9}
	B^{d_1d_2d_3}_{l_1l_2l_3} =&  \sum_c A^{d_1c} A^{d_2c} A^{d_3c} B^{c}_{l_1l_2l_3}.
\end{align}
This formalism allows us to also easily include primordial non-Gaussianity in our model. In order to do this we can add the bispectrum of the CMB, parametrized by $f_{\text{NL}}$. Under the same set of assumpions of the framework of SMICA defined in section~\ref{sec:framework} we can then write the full model bispectrum as
\begin{align}
	\label{eq:like:9b}
	B^{d_1d_2d_3}_{l_1l_2l_3} =& A^{d_1 \ \text{CMB}}A^{d_3 \ \text{CMB}}A^{d_4 \ \text{CMB} }f^{\text{th}}_{\text{NL}}B^{\text{th}}_{l_1l_2l_3} \nonumber\\
    & +A^{d_1 \ \text{Dust}}A^{d_3 \ \text{Dust}}A^{d_4 \ \text{Dust} }B^{\text{Dust}}_{l_1l_2l_3}\nonumber\\
    & +A^{d_1 \ \text{Sync}}A^{d_3 \ \text{Sync}}A^{d_4 \ \text{Sync} }B^{\text{Sync}}_{l_1l_2l_3},
\end{align}
where $A^{d_1 \ \text{CMB}}=1$ at all the frequencies analyzed, $B^{\text{th}}_{l_1l_2l_3}$ is the model dependent primordial bispectrum that we want to study (such as the local template).

\subsection{The binned bispectrum}
\label{sec:mdmcbb:bin}
In light of the complexity of the bispectrum, many diferent estimators have been developed in order to estimate the CMB primordial non-Gaussianities (PNG). Three main estimators were used in the \textit{Planck} papers~\cite{Planck_2013, Planck_2015, Planck_2018}: the KSW estimator~\cite{Komatsu_2005, Yadav_2007, Yadav_2008}, the binned estimator~\cite{Bucher_2010, Bucher_2016}, and the modal estimator~\cite{Fergusson_2010, Fergusson_2012, Fergusson_2014}. All of these estimators have in common the fact that they are implemented in the traditional pipeline analysis of the CMB after the component separation step. Indeed, all bispectrum analyses (both for the CMB and the foregrounds) have been performed on separated component maps, hence, at the end of the pipeline, after many different gaussianity assumptions had already been implemented, independently for each component. Given the multi-frequency multi-component framework implemented in SMICA and the extension to non-Gaussianity we have developed for it, we can move the bispectrum analysis to the same level as component separation, working with observed maps per frequency channel with the objective of doing a joint estimation of the bispectrum of each component. This approach not only shifts the analysis of non-Gaussianity to the component separation step, but also incorporates the foreground separation uncertainties directly in the estimation, as it combines data in an optimal way accounting for both the power spectrum and bispectrum of the components, unlike the standard approach to non-Gaussianity.

We adopted the binned bispectrum approximation. As explained in section~\ref{sec:nonG}, the bispectrum is essentially divided into two: the even $l_1+l_2+l_3$ and odd $l_1+l_2+l_3$ parts, the first one being real and symmetric and the second imaginary and anti-symmetric. It is therefore a natural extension for the binned bispectrum estimation to define two different bispectra in the two cases. In this section, as we did before, we still consider a single polarization (T, E or B) but it can be generalized to multiple polarization cross-bispectra with an extra index. Thus, the bispectrum is divided into:
\begin{itemize}
    \item Even part: the bispectrum is redefined as~\cite{Bucher_2016}
    \begin{align}
    \label{eq:bin:1}
        B^e_{l_1l_2l_3} = h^{000}_{l_1l_2l_3} B_{l_1l_2l_3} = h^{000}_{l_1l_2l_3} \sum_{m_i} \begin{pmatrix}
            l_1 & l_2 & l_3 \\ m_1 & m_2 & m_3
        \end{pmatrix} a_{l_1m_1}a_{l_2m_2}a_{l_3m_3},
    \end{align}
    where
    \begin{align}
    \label{eq:bin:2}
        h^{000}_{l_1l_2l_3}= \sqrt{\frac{(2l_1+1)(2l_2+1)(2l_3+1)}{4\pi}}\begin{pmatrix}
            l_1 & l_2 & l_3 \\ 0 & 0 & 0
        \end{pmatrix},
    \end{align}
    which is non-zero only for $l_1+l_2+l_3$ even. This allows for a very intuitive definiton of the observed even bispectrum via the filtered maps $M_l(\Omega)$
    \begin{align}
    \label{eq:bin:3}
        M_l(\Omega) = \sum_m \widehat{a}_{lm} \mathcal{Y}_{lm}(\Omega),
    \end{align}
    since 
    \begin{align}
    \label{eq:bin:4}
        &\int d\Omega \ M_{l_1}(\Omega)M_{l_2}(\Omega)M_{l_3}(\Omega) \nonumber\\
        =& \sum_{m_i} \widehat{a}_{l_1m_1}\widehat{a}_{l_2m_2}\widehat{a}_{l_3m_3}\int d\Omega \ \mathcal{Y}_{l_1m_1}(\Omega)\mathcal{Y}_{l_2m_2}(\Omega)\mathcal{Y}_{l_3m_3}(\Omega)\nonumber\\
        =& \sum_{m_i} h^{000}_{l_1l_2l_3}\begin{pmatrix}
            l_1 & l_2 & l_3 \\ m_1 & m_2 & m_3
        \end{pmatrix}\widehat{a}_{l_1m_1}\widehat{a}_{l_2m_2}\widehat{a}_{l_3m_3}=\widehat{B}^e_{l_1l_2l_3},
    \end{align}
    due to the Gaunt integral. The binned even bispectrum is then simply defined as 
    \begin{align}
    \label{eq:bin:5}
        B^e_{i_1i_2i_3} =& \sum_{l_i \in \Delta_i}  B^e_{l_1l_2l_3},
    \end{align}
    where $\Delta_i$ represents the $i$-th bin. The observed even binned bispectrum is computed via the binned filtered maps
    $M_i(\Omega)=\sum_{l\in \Delta_i} M_l(\Omega)$ 
    \begin{align}
    \label{eq:bin:6}
        \widehat{B}^e_{i_1i_2i_3}=&\int d\Omega \ M_{i_1}(\Omega)M_{i_2}(\Omega)M_{i_3}(\Omega).
    \end{align}
    The variance of the even binned bispectrum is
    \begin{align}
    \label{eq:bin:7}
        \mathrm{Var}(B^e)_{i_1i_2i_3} =& \ g_{i_1i_2i_3}\sum_{l_i \in \Delta_i}(h^{000}_{l_1l_2l_3})^2 C_{l_1}C_{l_2}C_{l_3},
    \end{align}
    where $g_{i_1i_2i_3}$ is $6$ if all indexes are equal, $2$ if only two of them are equal and $1$ if they are all different. Everything can be generalized to frequency space as 
    \begin{align}
    \label{eq:bin:8}
        B^{e, d_1d_2d_3}_{l_1l_2l_3} =& \sum_{l_i \in \Delta_i}\sum_{m_i} h^{000}_{l_1l_2l_3} \begin{pmatrix}
            l_1 & l_2 & l_3 \\ m_1 & m_2 & m_3
        \end{pmatrix}a^{d_1}_{l_1m_1}a^{d_2}_{l_2m_2}a^{d_3}_{l_3m_3},\\
        \left(\mathrm{Var}(B^e)\right)^{d_1d_2d_3,d'_1d'_2d'_3}_{i_1i_2i_3} =& g_{i_1i_2i_3}\sum_{l_i \in \Delta_i}(h^{000}_{l_1l_2l_3})^2 C^{d_1d'_1}_{l_1}C^{d_2d'_2}_{l_2}C^{d_3d'_3}_{l_3}.
    \end{align}

    \item Odd part: in this case we need to define a new $h$ factor~\cite{Coulton_2019}, using the properties of the spin-weighted spherical harmonics $_s\mathcal{Y}_{lm}$
    \begin{align}
    \label{eq:bin:9}
        h^{2 -1 -1}_{l_1l_2l_3} =& \sqrt{\frac{(2l_1+1)(2l_2+1)(2l_3+1)}{4\pi}} \begin{pmatrix}
            l_1 & l_2 & l_3 \\ 2 & -1 & -1 
        \end{pmatrix}(1-(-1)^{l_1+l_2+l_3}),
    \end{align}
    which is non-zero only for $l_1+l_2+l_3$ odd. With this we can define the odd part of the bispectrum as 
    \begin{align}
    \label{eq:bin:10}
        B^o_{l_1l_2l_3} = \frac{1}{6}\Big( h^{2 -1 -1}_{l_1l_2l_3}+ h^{2 -1 -1}_{l_3l_1l_2}  + h^{2 -1 -1}_{l_2l_3l_1}\Big)\sum_{m_i}\begin{pmatrix}
            l_1 & l_2 & l_3 \\ m_1 & m_2 & m_3 
        \end{pmatrix}  a_{l_1m_1}a_{l_2m_2}a_{l_3m_3}.
    \end{align}
    Similarly to the even case, we can connect this to the observed odd bispectrum via the filtered spin-weighted maps
    \begin{align}
    \label{eq:bin:11}
        _{s}M_l(\Omega) = \sum_{m} \widehat{a}_{lm} \ _{s}\mathcal{Y}_{lm}(\Omega)
    \end{align}
    since
    \begin{align}
    \label{eq:bin:12}
        &\frac{1}{6}\int d\Omega \ \Big( \   _{-2}M_{l_1}(\Omega) _{1}M_{l_2}(\Omega) _{1}M_{l_3}(\Omega)- \ _{2}M_{l_1}(\Omega) _{-1}M_{l_2}(\Omega) _{-1}M_{l_3}(\Omega)\nonumber\\
        &+\  _{-2}M_{l_3}(\Omega)_{1}M_{l_1}(\Omega)_{1}M_{l_2}(\Omega)- \ _{2}M_{l_3}(\Omega)_{-1}M_{l_1}(\Omega)_{-1}M_{l_3}(\Omega)\nonumber\\
        &+\  _{-2}M_{l_2}(\Omega)_{1}M_{l_3}(\Omega)_{1}M_{l_1}(\Omega)- \ _{2}M_{l_2}(\Omega)_{-1}M_{l_3}(\Omega)_{-1}M_{l_1}(\Omega) \Big)\nonumber\\
        =&\frac{1}{6}\Big( h^{2 -1 -1}_{l_1l_2l_3}+ h^{2 -1 -1}_{l_3l_1l_2}  + h^{2 -1 -1}_{l_2l_3l_1}\Big)\sum_{m_i}\begin{pmatrix}
            l_1 & l_2 & l_3 \\ m_1 & m_2 & m_3 
        \end{pmatrix} \widehat{a}_{l_1m_1}\widehat{a}_{l_2m_2}\widehat{a}_{l_3m_3}=\widehat{B}^o_{l_1l_2l_3},
    \end{align}
    due to the properties of the spin-weighted spherical harmonics. This way, we ensure to keep only the odd triplets inside our bins. Indeed if we define
    \begin{align}
    \label{eq:bin:13}
        B^o_{i_1i_2i_3} =& \sum_{l_i \in \Delta_i} B^o_{l_1l_2l_3},
    \end{align}
    we can compute the odd observed binned bispectrum via the binned spin-weighted filtered maps $_{s}M_{i}(\Omega)=\sum_{l\in\Delta_i} \ _{s}M_{l}(\Omega)$ as
    \begin{align}
    \label{eq:bin:14}
        \widehat{B}^o_{i_1i_2i_3}=&\frac{1}{6}\int d\Omega \ \Big( \   _{-2}M_{i_1}(\Omega) _{1}M_{i_2}(\Omega)_{1}M_{i_3}(\Omega)- \ _{2}M_{i_1}(\Omega)_{-1}M_{i_2}(\Omega)_{-1}M_{i_3}(\Omega)\nonumber\\
        &+\  _{-2}M_{i_3}(\Omega)_{1}M_{i_1}(\Omega)_{1}M_{i_2}(\Omega)- \ _{2}M_{i_3}(\Omega)_{-1}M_{i_1}(\Omega)_{-1}M_{i_3}(\Omega)\nonumber\\
        &+\  _{-2}M_{i_2}(\Omega)_{1}M_{i_3}(\Omega)_{1}M_{i_1}(\Omega)- \ _{2}M_{i_2}(\Omega)_{-1}M_{i_3}(\Omega)_{-1}M_{i_1}(\Omega) \Big).
    \end{align}
    The variance of the odd binned bispectrum is
    \begin{align}
    \label{eq:bin:15}
        \mathrm{Var}(B^o)_{i_1i_2i_3} =& g_{i_1i_2i_3}\sum_{l_i\in\Delta_i}\frac{(1-(-1)^{l_1+l_2+l_3})^2}{6^2}\left[ \begin{pmatrix}
            l_1 & l_2 & l_3 \\ 2 & -1 & -1 
        \end{pmatrix} + \begin{pmatrix}
            l_3 & l_1 & l_2 \\ 2 & -1 & -1 
        \end{pmatrix} + \begin{pmatrix}
            l_2 & l_3 & l_1 \\ 2 & -1 & -1 
        \end{pmatrix}\right]^2 \nonumber\\
        &\times C_{l_1}C_{l_2}C_{l_3}.
    \end{align}
    All of this can be generalized to frequency space in the usual way.
\end{itemize}
With these two definitions, we will now estimate the bispectra with the likelihood in eq.~\eqref{eq:like:8} independently for the even and odd parts.

\subsection{B-mode polarizations results}
\label{sec:res2:B}
The scope of this analysis was to reconstruct the bispectra of the foregrounds, highly dependent on the mask, therefore we have worked with an unmasked sky using the same set of 400 simulations described in section~\ref{sec:sim}: gaussian CMB, dust (d0), synchrotron (s0) and gaussian noise. We maximized the likelihood in eq.~\eqref{eq:like:8} with respect to the dust and synchrotron bispectra, both for the even and odd parts. The binning choice has been the following:
\begin{align}
    \label{eq:res2:B1}
    \Delta_i \in \{&[2,5],[6,10],[11,15],[16,20],[21,30],[31,40],\nonumber\\
    &[41,50],[51,60],[61,70],[71,80],[81,90],[91,100]\}.
\end{align}

\begin{figure}[!htbp]
\hspace*{-1cm} 
\centering 
\includegraphics[scale=0.5]{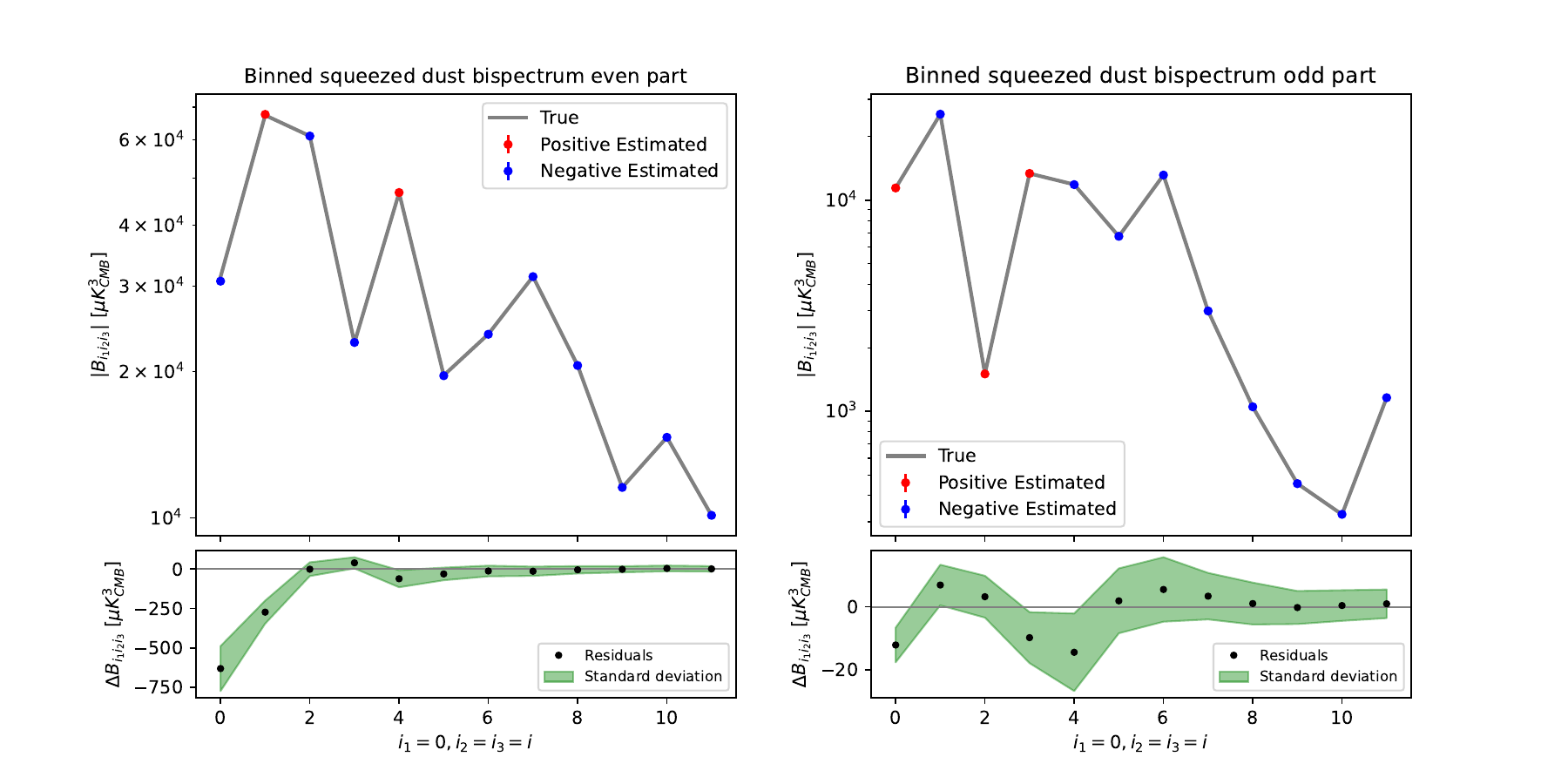}
\caption{\label{fig:res2:d_loc_b} Upper plots: the "true" dust input map bispectrum for the B-mode and the estimated dust bispectrum averaged over 400 simulations, both for the  even (left) and odd (right) sectors in the squeezed configuration $i_1=0$ and $i_2=i_3=i$, as a function of $i$. We indicate positive values as red points and negative ones as blue points for the estimated bispectrum.
Lower plots: the comparison between the residuals and the standard deviation over the 400 simulations.}
\end{figure}

\begin{figure}[!htbp]
\hspace*{-1cm} 
\centering 
\includegraphics[scale=0.5]{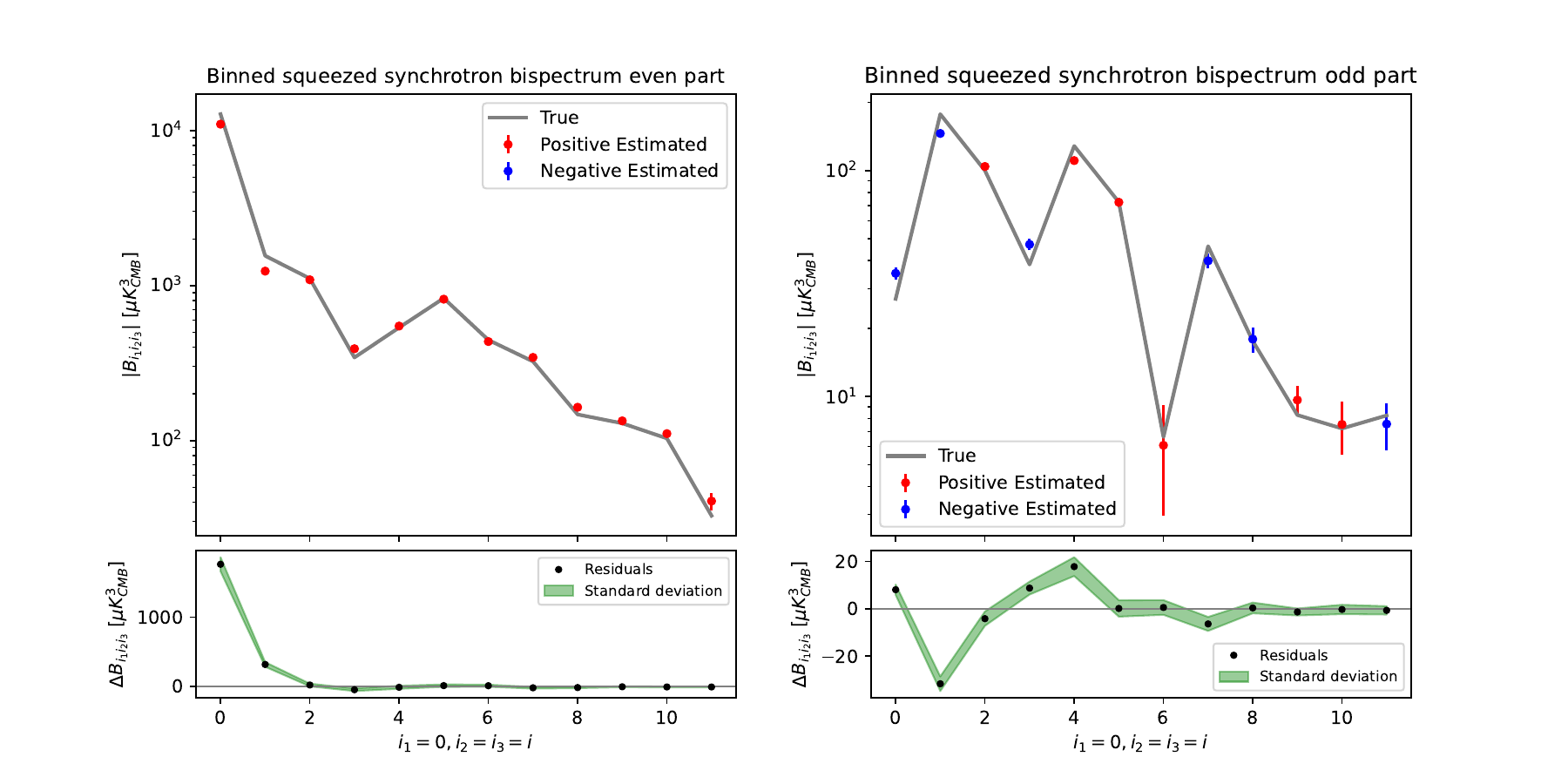}
\caption{\label{fig:res2:s_loc_b} Same as figure~\ref{fig:res2:d_loc_b} but for synchrotron.}
\end{figure}

\begin{figure}[!htbp]
\hspace*{-1cm} 
\centering 
\includegraphics[scale=0.5]{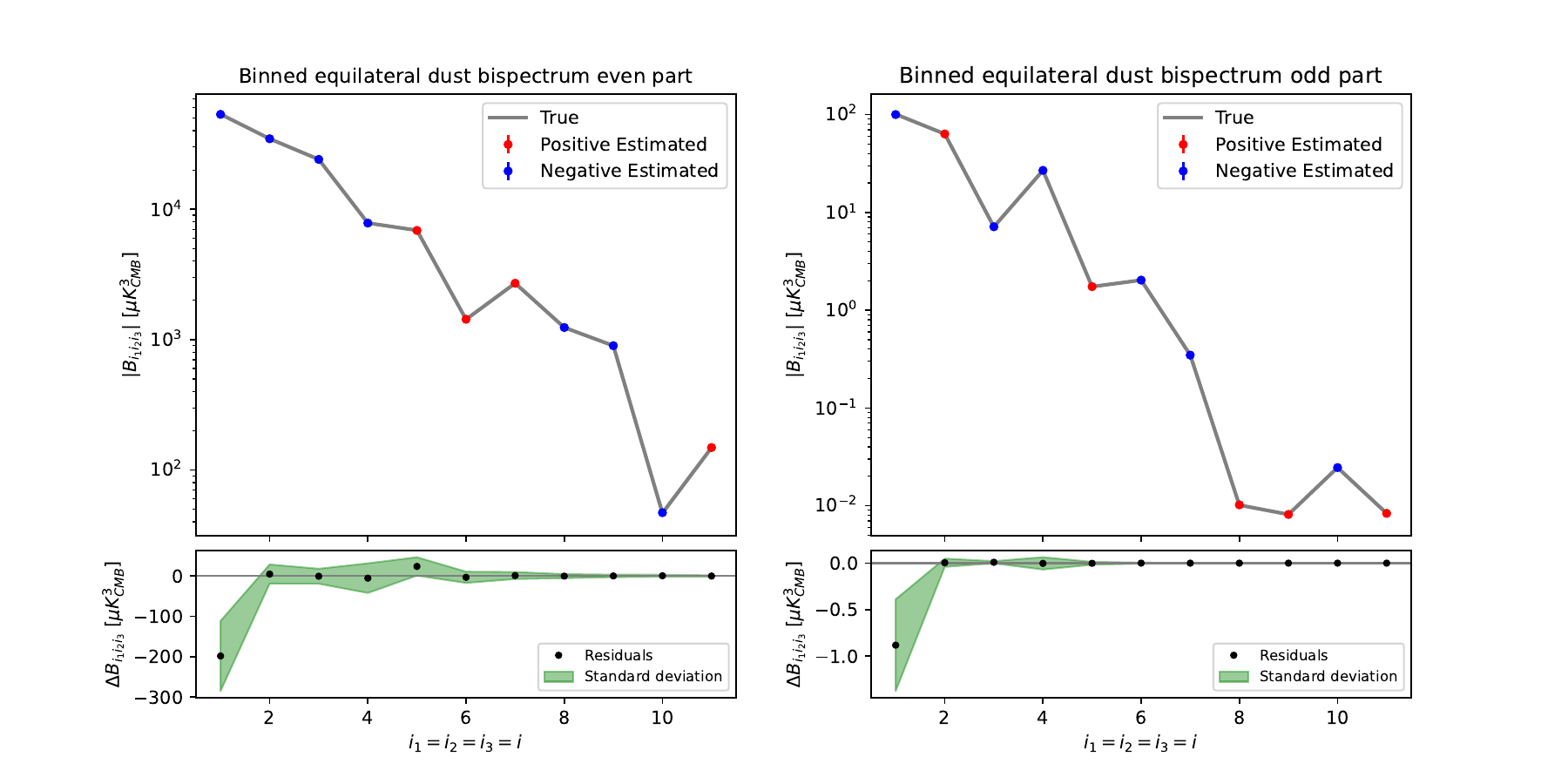}
\caption{\label{fig:res2:d_eq_b} Same as figure~\ref{fig:res2:d_loc_b} but in the equilateral configuration $i_1=i_2=i_3=i$ as a function of $i$.}
\end{figure}

\begin{figure}[!htbp]
\hspace*{-1cm} 
\centering 
\includegraphics[scale=0.5]{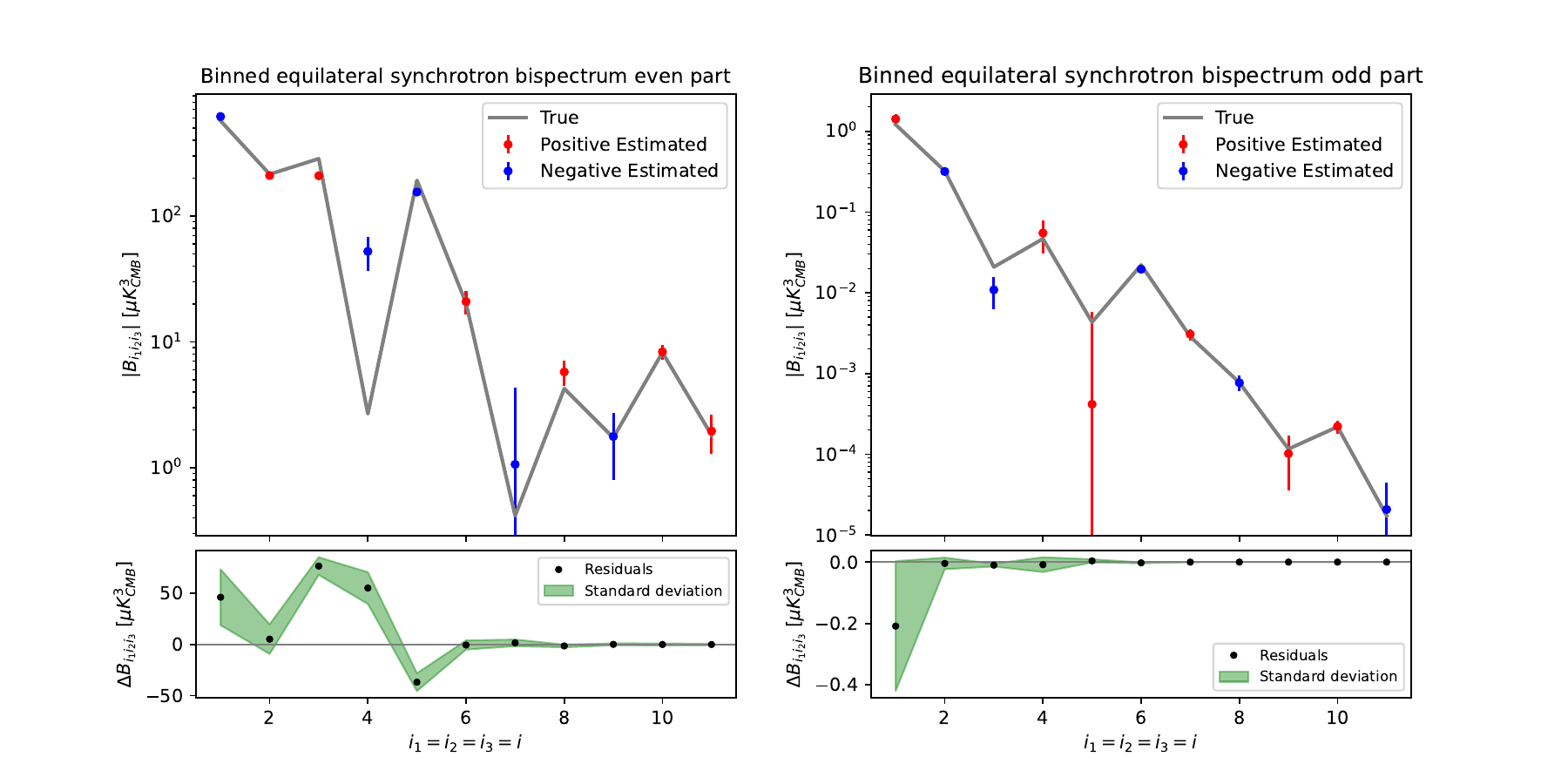}
\caption{\label{fig:res2:s_eq_b} Same as figure~\ref{fig:res2:d_loc_b} but for synchrotron and in the equilateral configuration $i_1=i_2=i_3=i$ as a function of $i$.}
\end{figure}

We divide the results into two sectors: the squeezed configuration, which corresponds to $i_1=0$ and $i_2=i_3$, and the equilateral configuration, which instead corresponds to $i_1=i_2=i_3$. In figure~\ref{fig:res2:d_loc_b} and figure~\ref{fig:res2:s_loc_b} we show the results obtained for the first case and in figure~\ref{fig:res2:d_eq_b} and figure~\ref{fig:res2:s_eq_b} the results obtained for the latter. We compare our estimated bispectrum averaged over the 400 simulations with the bispectrum computed from the original input maps (d0 and s0). We also plot the corresponding residual, defined as the difference between the estimated bispectrum and the original one, with the $1\sigma$ confidence region obtained from the simulations. As discussed in section~\ref{subsec:datamodel}, the squeezed configuration is the most significant for both foregrounds, as it is where the bispectra have the strongest signal-to-noise ratio. Our approach proved to be effective in recovering the correct sign and amplitude of the bispectrum for most triplets $i_1\leq i_2\leq i_3$, with the dust bispectrum being the better constrained by the set of frequencies studied. Our estimation is limited by the weak signal-to-noise ratio of the bispectrum, improved by the choice of binning, which is a choice based on the assumption of smooth bispectra over the bins, which is an important approximation in the case of the two foregrounds.

\FloatBarrier

\subsection{E-mode polarizations results}
\label{sec:res2:E}
Similarly to section~\ref{sec:res2:B} we have performed the same analysis on the simulations for the E-mode polarization: the objective was to include in the model also primordial non-Gaussianity in the CMB. One of the most studied types of bispectrum non-Gaussianity in the CMB is the local bispectrum~\cite{Gangui_1994}, mostly related to multiple-field inflation, where curvature perturbations can evolve on superhorizon scales~\cite{Rigopoulos_2007, Byrnes_2010}. It is defined by~\cite{Komatsu_2001} 
\begin{align}
    \label{eq:res2:E1}
    (B^{\text{loc}})^{p_1p_2p_3}_{l_1l_2l_3} =& (h^{000}_{l_1l_2l_3})^2 \left( \frac{2}{\pi}\right)^3 \int_0^{+\infty} k_1^2dk_1 \int_0^{+\infty}k_2^2dk_2  \int_0^{+\infty}k_3^2dk_3 \Big[ g^{p_1}_{l_1}(k_1)g^{p_2}_{l_2}(k_2) g^{p_3}_{l_3}(k_3)B^{\text{loc}}(k_1,k_2,k_3)\nonumber\\
    &\times \int_0^{+\infty} r^2 dr \  j_{l_1}(k_1r)j_{l_2}(k_2r)j_{l_3}(k_3r)\Big],
\end{align}
where $j_l$ are spherical Bessel functions, $g^p_l$ are the radiation transfer functions for the polarization $p$ and $B^{\text{loc}}$ is the primordial bispectrum of the adiabatic curvature perturbation $\zeta$ defined as 
\begin{align}
    \label{eq:res2:E1.1}
    B^{\text{loc}}(k_1,k_2,k_3) = -\frac{6}{5}\Big[ P(k_1)P(k_2) + \ \text{2 perms}\Big],
\end{align}
where $P(k) = \frac{2\pi^2}{k^3}P_\zeta(k)$ is the rescaled primordial adiabatic power spectrum.
In our case, of E-mode polarization only, the  bispectrum of the CMB is parametrized as 
\begin{align}
    \label{eq:res2:E2}
    B^{\text{CMB}}_{l_1l_2l_3} = f^{\text{loc}}_{\text{NL}} B^{\text{loc}}_{l_1l_2l_3},
\end{align}
where $\text{loc}$ means that it refers to the local type of non-Gaussianity. The parameter $f_{\text{NL}}$ can then be estimated, given an observed bispectrum $\widehat{B}$ and any theoretical bispectrum template $B^{\text{th}}$, using
\begin{align}
    \label{eq:res2:E3}
    f^{th}_{\text{NL}} = \frac{\langle \widehat{B},B^{\text{th}}\rangle}{\langle B^{\text{th}},B^{\text{th}}\rangle},
\end{align}
where the inner product is the one in eq.~\eqref{eq:edge:4} for a single map
\begin{align}
    \label{eq:res2:E4}
    \langle B^I,B^{J}\rangle= \sum_{l_1\leq l_2\leq l_3} B^I_{l_1l_2l_3}\mathrm{Var}(\textbf{B})^{-1}_{l_1l_2l_3}B^J_{l_1l_2l_3}.
\end{align}

In order to compare the results of our fit we have also performed the "standard" non-gaussian analysis, reconstructing via the Generalised Least Square (GLS) filter the CMB and noise maps after component separation  and applying the usual binned bispectrum estimator on them. Under gaussian statistics and without other prior assumptions on the astrophysical components, given the linear multi-frequency and multi-component model in eq.~\eqref{eq:frame:1}, one can use a generalized least-square solution in harmonic space~\cite{Tegmark_1997, Delabrouille_2003}
\begin{align}
    \label{eq:res2:E6}
    \widehat{S} = (A^T N^{-1}A)^{-1} AN^{-1} Y,
\end{align}
where $A$ is the mixing matrix and $N$ is the noise covariance matrix, both estimated by SMICA. $Y$ instead represents the observed frequency maps in harmonic space, hence is a set of $N_{\text{freq}}$ $a_{lm}$'s. The GLS solution is sub-optimal, but it allows for a fast comparison between the different approaches.

\begin{figure}[!htbp]
\centering 
\includegraphics[scale=0.7]{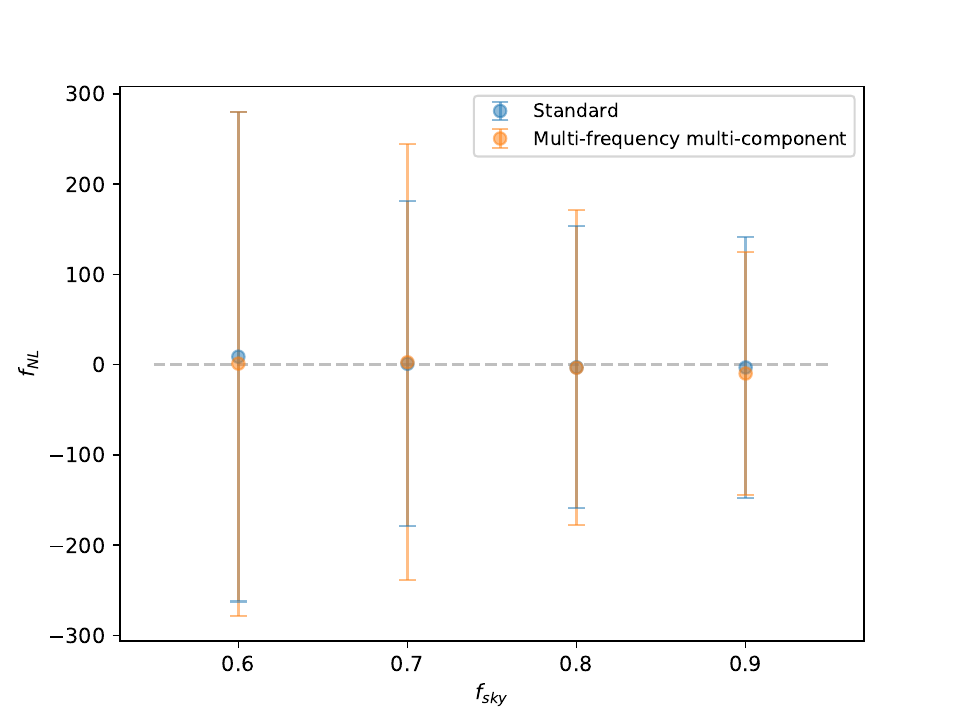}
\caption{\label{fig:res2:fnl} Comparison of the results for $f_{\text{NL}}$ and their error bars between the standard binned bispectrum estimator and the multi-frequency multi-component binned bispectrum estimator for the 400 simulations for masks with a unmasked sky fraction of 60\%, 70\%, 80\% and 90\%.}
\end{figure}

In figure~\ref{fig:res2:fnl} we show the results obtained with our multi-frequency multi-component binned bispectrum estimator and with the standard binned bispectrum estimator for different masks. In the two cases, we recover consistent results. This was expected as we are working with a simplified foreground model, in which case the GLS filter efficiently removes foreground residuals from the CMB map. With a more complex sky, our approach is expected to be more robust and more efficient in detecting deviations from the idealized model, as the foregrounds are already included in the analysis. The large error bars in figure~\ref{fig:res2:fnl} might surprise since the current constraint on $f_{\text{NL}}$ for the local shape is $\sigma(f_{\text{NL}})=5$ given by the analysis of the Planck data \cite{ Jung_2025}, but it is important to note that that limit was obtained via a combined temperature and E-mode analysis going up to $\ell_{\text{max}}=2500$, while here we have E only and $\ell_{\text{max}}=100$. Moreover, in this analysis we have not included the linear correction term, which is required to obtain optimal error bars if isotropy is broken, as is the case here because of the mask and anisotropic foregrounds.

\begin{figure}[!htbp]
\hspace*{-1cm} 
\centering 
\includegraphics[scale=0.5]{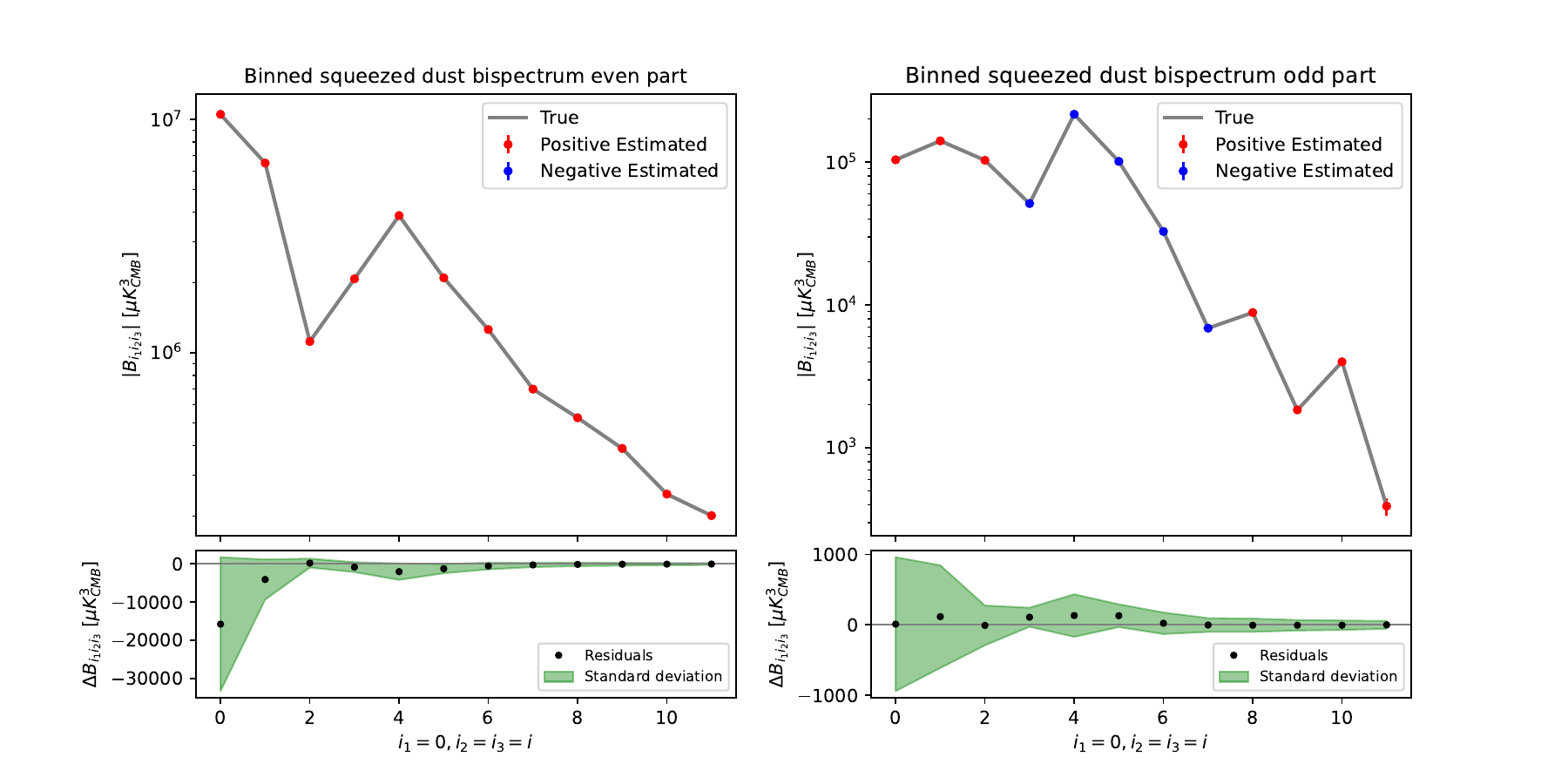}
\caption{\label{fig:res2:d_loc_e} Upper plots: the "true" dust input map bispectrum for the E-mode and the estimated dust bispectrum averaged over 400 simulations, both for the  even (left) and odd (right) sectors in the squeezed configuration $i_1=0$ and $i_2=i_3=i$, as a function of $i$. We indicate positive values as red points and negative ones as blue points for the estimated bispectrum.
Lower plots: the comparison between the residuals and the standard deviation over the 400 simulations.}
\end{figure}

\begin{figure}[!htbp]
\hspace*{-1cm} 
\centering 
\includegraphics[scale=0.5]{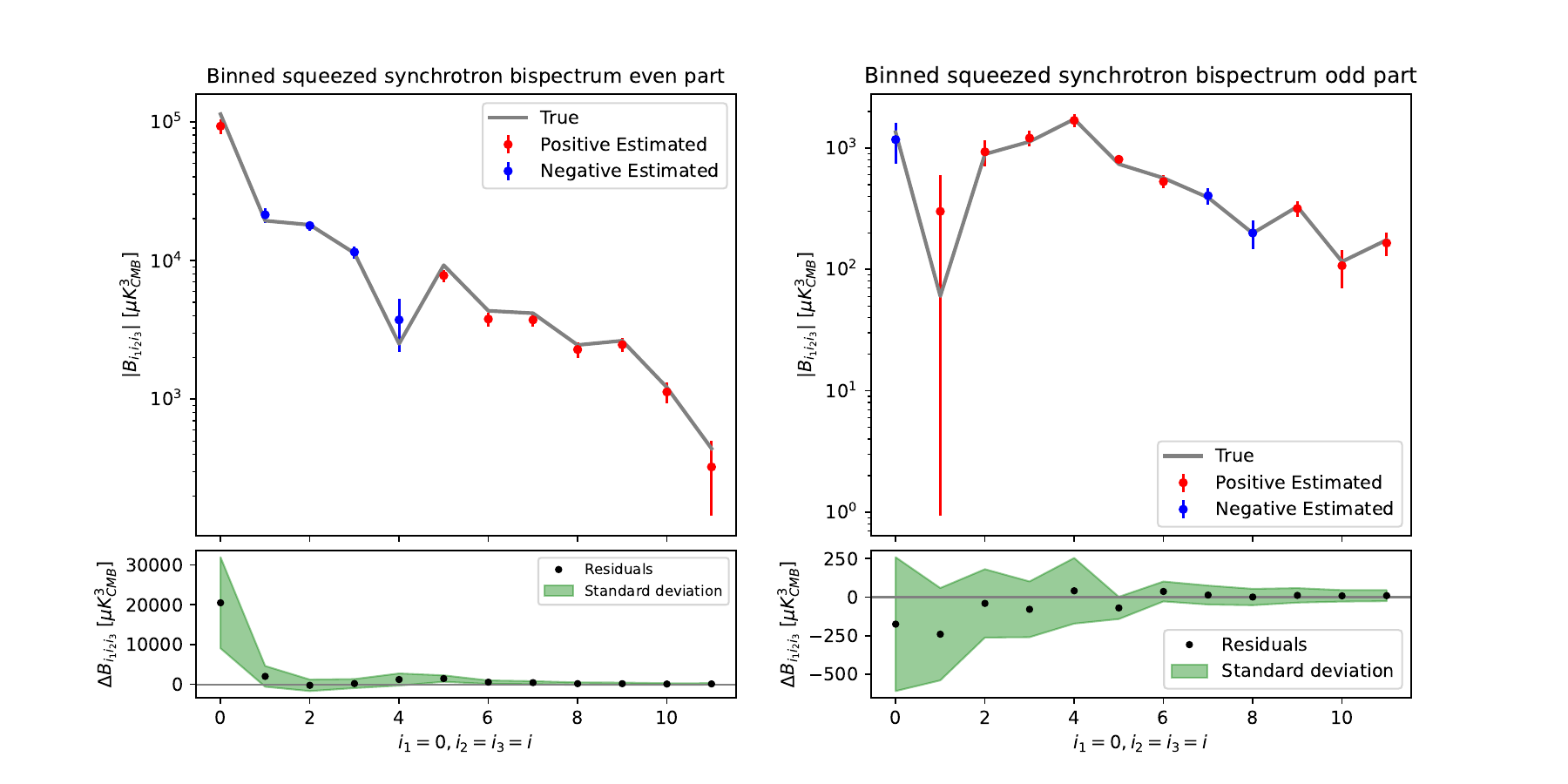}
\caption{\label{fig:res2:s_loc_e} Same as figure~\ref{fig:res2:d_loc_e} but for synchrotron.}
\end{figure}

\begin{figure}[!htbp]
\hspace*{-1cm} 
\centering 
\includegraphics[scale=0.5]{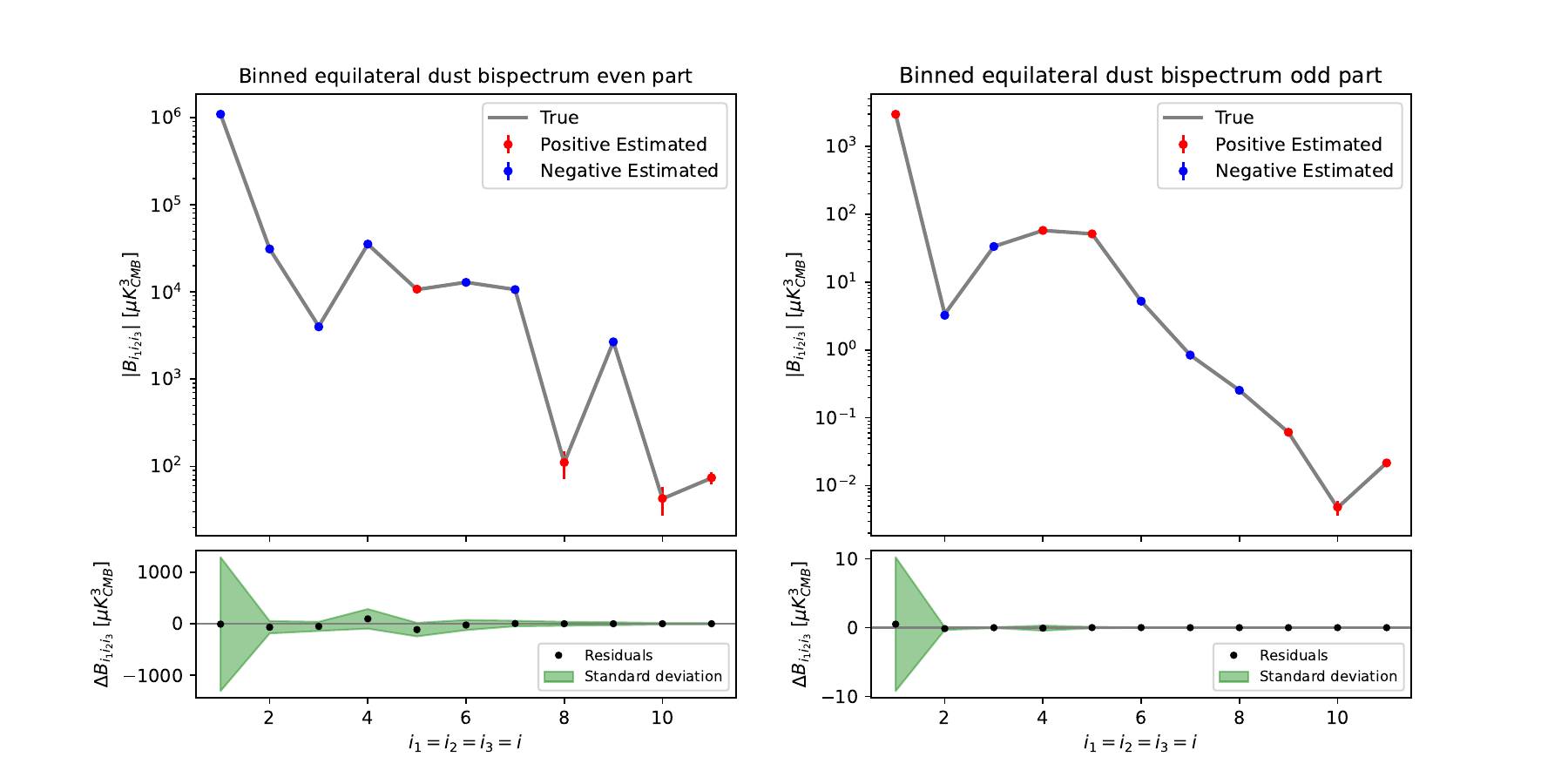}
\caption{\label{fig:res2:d_eq_e} Same as figure~\ref{fig:res2:d_loc_e} but in the equilateral configuration $i_1=i_2=i_3=i$ as a function of $i$.}
\end{figure}

\begin{figure}[!htbp]
\hspace*{-1cm} 
\centering 
\includegraphics[scale=0.5]{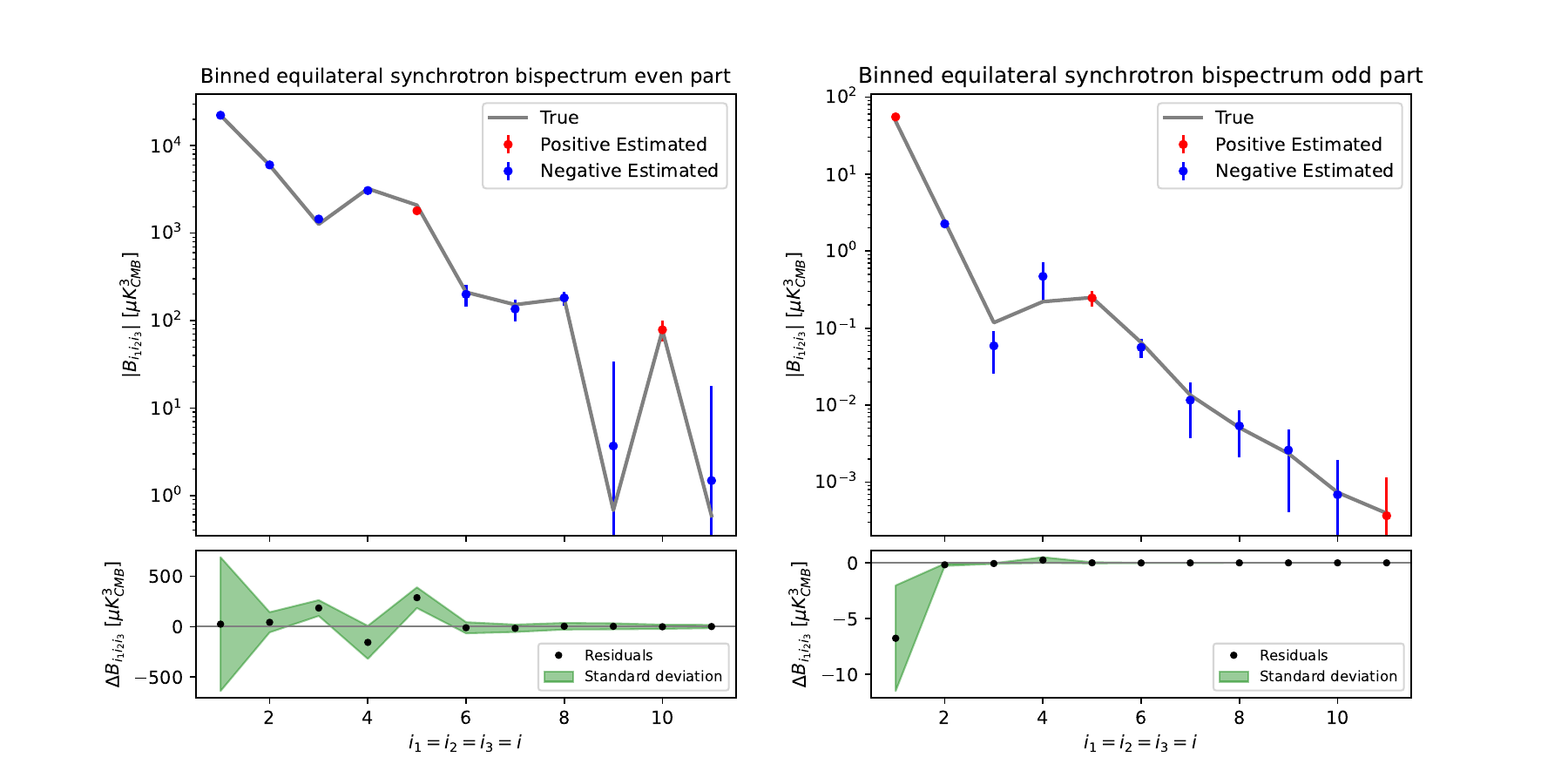}
\caption{\label{fig:res2:s_eq_e} Same as figure~\ref{fig:res2:d_loc_e} but for synchrotron and in the equilateral configuration $i_1=i_2=i_3=i$ as a function of $i$.}
\end{figure}

As in section~\ref{sec:res2:B} we show the results obtained for the foregrounds in the squeezed configuration in figure~\ref{fig:res2:d_loc_e} and figure~\ref{fig:res2:s_loc_e} and the results in the equilateral configuration in figure~\ref{fig:res2:d_eq_e} and figure~\ref{fig:res2:s_eq_e}. In the E-mode polarization case the significance of the squeezed configuration over the equilateral one is much more evident for the two foregrounds. Indeed, the squeezed configuration is recovered without any bias for both dust and synchrotron, while the equilateral bispectrum of the foregrounds has a lower amplitude that tends to zero. It is interesting to compare our results with those obtained in~\cite{Coulton_2019} for the SPASS $2.3$~GHz, Planck PR3 $30-857$~GHz and IRAS $3$~THz maps. The authors find no evidence of polarized synchrotron bispectra, consistent with our findings. Moreover, they report no evidence for cross-bispectra between dust and synchrotron emission. This assumption underlies our method and appears to be valid, as we recover the foreground bispectra without bias. Finally, they also agree that the dust bispectrum exhibits its strongest signal in the squeezed configuration.

With this method, we have effectively shifted the standard analysis of non-Gaussianity to frequency channel maps, dealing simultaneously with multiple components. This will allow us to constrain primordial non-Gaussianity without any gaussian assumption on the distribution of the primordial fluctuations, while coherently removing the non-gaussian signal produced by the foregrounds. In the standard analysis of primordial non-Gaussianity there is no propagation of the error of the component separation step. However, our approach allows us to coherently incorporate it into the bispectrum analysis.

\FloatBarrier

\FloatBarrier

\section{Conclusions}
\label{sec:conc}
Given the new CMB experiments coming online in the next 10 years, it is important to improve our existing component separation methods to the level required by those experiments. The objective of this work was to prove that it is possible to merge the analysis of non-Gaussianity both for the foregrounds and for the CMB with the component separation step. We achieved this using two approaches: first, by including higher-order statistics into the component separation process and second, by developing a bispectrum estimator capable of recovering the correct bispectrum for different components at the same time, working with frequency maps.  The component separation method that we used is SMICA, which is non-parametric and works in the harmonic domain.

We included higher-order statistics via the Multivariate Edgeworth Expansion (MEE), which is an expansion in terms of cumulants around a gaussian distribution capable of approximating any PDF. We truncated this expansion at the first non-gaussian order with the purpose of including the bispectrum. The result we found is that there is no improvement in the power spectrum and mixing matrix estimation by including information of the $3$-point correlator in the component separation process. The physical reasons behind this result might be the lack of a well-motivated theoretical model for the bispectrum of the foregrounds or the fact that the bispectrum alone is not enough to capture the complexity of the foregrounds. We have also demonstrated that the MEE approximation is not well suited for a joint estimation of the bispectra together with the power spectra and mixing matrix components. 

Based on these results, we developed an independent bispectrum estimator: since a combined power spectrum and bispectrum analysis has shown no improvement for the power spectrum estimation, we developed a multi-frequency likelihood for the binned bispectrum and performed a conditioned estimation of the bispectrum with the spectral parameters obtained independently by SMICA. With this estimator, we were able to recover the correct bispectra of the two foregrounds included in our simulations, dust and synchrotron, while also recovering coherent constraints on local primordial non-Gaussianity with respect to the standard post-component separation analysis. However, our approach allows for a better handling of the foregrounds, in particular in the case of more complicated foreground models where we expect our method to outperform the standard approach.

We have tested our method on simulations based on the LiteBIRD mission configuration~\cite{LB_2025} working with E and B polarizations independently. In a future work, we intend to test our approach with respect to the standard one via gaussian simulations of multi-component multi-frequency sky observations, for which the assumptions of SMICA would be perfectly accurate, in order to estimate if the standard analysis is underestimating the error bars of PNG. While waiting for the improved PNG constraints that new data from future CMB experiments will deliver, we intend to generalize our new estimator to multiple polarizations and apply it to the 9 frequency channels of Planck, which requires to generalize gaussian SMICA to work jointly on multiple polarizations, an ongoing project. In order to do this Planck analysis, we will also implement the linear correction term. It is simple to implement in our formalism but it is very time-consuming to run, which is why we omitted it in this paper. Finally, in this work we focused on polarized \textit{galactic} foregrounds, but our approach might prove particularly effective in reducing the bias on $f_{\mathrm{NL}}$ introduced by \textit{extra-galactic} foregrounds, such as the thermal Sunyaev–Zel’dovich effect and the Cosmic Infrared Background. The main challenge they present is that these effects contaminate smaller angular scales, posing a greater numerical challenge for our analysis and thus requiring a broader binning scheme. However, unlike their galactic counterparts, these contaminants are statistically isotropic and their bispectra can be modeled with reasonable accuracy using analytical and simulation-based methods. If left uncorrected, these foregrounds can introduce significant biases in high-resolution CMB-based $f_{\mathrm{NL}}$ estimates~\cite{Hill_2018}, but our formalism could tackle this problem directly.

\acknowledgments
We would like to thank for many interesting discussions and inputs regarding non-Gaussianity and component separation methods: Clément Leloup, Wuhyun Sohn, Léo Vacher,  Josquin Errard, Michele Liguori and Benjamin Beringue. The authors acknowledge the use of the healpy and HEALPix packages~\cite{Zonca_2019, Gorski_2005}. We gratefully acknowledge the IN2P3 Computer Centre (\url{https://cc.in2p3.fr}) for providing the computing resources and services needed for the analysis with the binned bispectrum estimator. Project funded by the CMB-Inflate program: the European Union’s Horizon 2020 research and innovation program under the Marie Skłodowska-Curie grant agreement No 101007633 (\url{https://sites.google.com/view/cmb-inflate/home}).
This work has received support from France 2030 through the project named Académie Spatiale d’Île-de-France (https://academiespatiale.fr/) managed by the National Research Agency under bearing the reference ANR-23-CMAS-0041.

\appendix
\section{Multivariate Edgeworth expansion computations}
\label{app:MEE}
A gaussian statistically isotropic random spherical field is completely defined by the distribution of its spherical harmonic coefficients~\cite{Marinucci_2011,Lang_2015}. If we define the real vector $ x_i=(\mathrm{Re}(a^d_{lm>0}),a^d_{l0},\mathrm{Im}(a^d_{lm>0}))$, where the index $i$ represents $(d,l,m)$, its components are gaussianly distributed with variance $C_l$ for $a^d_{l0}$ and $\frac{1}{2}C_l$ for both $\mathrm{Re}(a^d_{lm>0})$ and $\mathrm{Im}(a^d_{lm>0})$. The distribution can be written as
\begin{align}
\label{eq:app:3}
    \mathrm{PDF}(\textbf{x}) =& \frac{1}{\sqrt{(2\pi)^{r}\det(\textbf{K})}} \exp\left(-\frac{1}{2}\sum_{i,i'} x_i \left(K^{-1}\right)^{i i'}x_{i'}\right)
\end{align}
where $r=N_{\text{freq}}\sum_{l=0}^{l_{\text{max}}} (2l+1)$ is the length of the vector $\textbf{x}$, and 
\begin{align}
\label{eq:app:4}
    K^{ii'} =& \delta^{ll'}\begin{pmatrix}
                \delta^{mm'}\frac{1}{2}C^{dd'}_l & 0 & 0 \\
                0 & C^{dd'}_l & 0 \\
                0 & 0 & \delta^{mm'}\frac{1}{2}C^{dd'}_l 
              \end{pmatrix} \nonumber \\ 
    \left(K^{-1}\right)^{ii'} =& \delta^{ll'}\begin{pmatrix}
                2\delta^{mm'} \left(C^{-1}_l\right)^{dd'} & 0 & 0 \\
                0 & \left(C^{-1}_l\right)^{dd'} & 0 \\
                0 & 0 & 2\delta^{mm'} \left(C^{-1}_l\right)^{dd'}
              \end{pmatrix},
\end{align}
where $\delta^{mm'}$ gives a block matrix structure only to the upper and lower part of the matrix as the middle part of the vector $\textbf{x}$ is $a^d_{l0}$ which has $m$ fixed to 0.

Now that we have a multivariate real gaussian distribution, we can apply the standard multivariate Edgeworth expansion to add the first-order non-gaussian term~\cite{Juszkiewicz_1995, Amendola_1996, Taylor_2001, Bartolo_2012,  Hall_2022}:
\begin{align}{2}
\label{eq:app:5a}
    \text{PDF}(\textbf{x}) =& \frac{1}{\sqrt{(2\pi)^k\det(\textbf{K})}} \exp{\left(-\frac{1}{2}\sum_{i,i'} x_i \left(K^{-1}\right)^{ii'}x_{i'}\right)}\Bigg[ 1+\frac{1}{6}\sum_{i,j,k,i',j',k'}\langle x_ix_jx_k\rangle\nonumber\\
    &\times \Bigg(\left(K^{-1}\right)^{ii'}\left(K^{-1}\right)^{jj'}\left(K^{-1}\right)^{kk'}x_{i'} x_{j'}x_{k'}\nonumber\\
    &\quad -\left(K^{-1}\right)^{ii'} \left(K^{-1}\right)^{jk}x_{i'} -\left(K^{-1}\right)^{jj'} \left(K^{-1}\right)^{ik}x_{j'}-\left(K^{-1}\right)^{kk'} \left(K^{-1}\right)^{ij}x_{k'}\Bigg)\Bigg]
\end{align}
It is possible to recover the distribution in terms of the complex $a_{lm}$'s. If we focus on the first term in the non-gaussian part and on the first element
\begin{align}
\label{eq:app:6}
    &\langle x_ix_jx_k\rangle \ \left(K^{-1}\right)^{ii'}\left(K^{-1}\right)^{jj'}\left(K^{-1}\right)^{kk'} \ x_{i'} x_{j'}x_{k'} \nonumber\\
    =& 2 \langle \mathrm{Re}(a^d_{lm>0})x_jx_k\rangle \ \left(C^{-1}_l\right)^{dd'}\left(K^{-1}\right)^{jj'}\left(K^{-1}\right)^{kk'} \ \mathrm{Re}(a^{d'}_{lm>0}) x_{j'}x_{k'} \nonumber\\
    & +2  \langle \mathrm{Im}(a^d_{lm>0})x_jx_k\rangle \ \left(C^{-1}_l\right)^{dd'}\left(K^{-1}\right)^{jj'}\left(K^{-1}\right)^{kk'} \ \mathrm{Im}(a^{d'}_{lm>0}) x_{j'}x_{k'} \nonumber\\
    & + \langle a^d_{l0}x_jx_k\rangle \ \left(C^{-1}_l\right)^{dd'}\left(K^{-1}\right)^{jj'}\left(K^{-1}\right)^{kk'} \ a^{d'}_{l0} x_{j'}x_{k'}\nonumber\\
    =&  \langle \mathrm{Re}(a^d_{lm>0})x_jx_k\rangle \ \left(C^{-1}_l\right)^{dd'}\left(K^{-1}\right)^{jj'}\left(K^{-1}\right)^{kk'} \ \mathrm{Re}(a^{d'}_{lm>0}) x_{j'}x_{k'} \nonumber\\
    & +  (-i)\langle \mathrm{Im}(a^d_{lm>0})x_jx_k\rangle \ \left(C^{-1}_l\right)^{dd'}\left(K^{-1}\right)^{jj'}\left(K^{-1}\right)^{kk'} \ i \mathrm{Im}(a^{d'}_{lm>0}) x_{j'}x_{k'} \nonumber\\
    &+ \langle \mathrm{Re}(a^d_{lm>0})x_jx_k\rangle \ \left(C^{-1}_l\right)^{dd'}\left(K^{-1}\right)^{jj'}\left(K^{-1}\right)^{kk'} \ \mathrm{Re}(a^{d'}_{lm>0}) x_{j'}x_{k'} \nonumber\\
    & +  i\langle \mathrm{Im}(a^d_{lm>0})x_jx_k\rangle \ \left(C^{-1}_l\right)^{dd'}\left(K^{-1}\right)^{jj'}\left(K^{-1}\right)^{kk'} \ (-i)\mathrm{Im}(a^{d'}_{lm>0}) x_{j'}x_{k'} \nonumber\\
    & + \langle a^d_{l0}x_jx_k\rangle \ \left(C^{-1}_l\right)^{dd'}\left(K^{-1}\right)^{jj'}\left(K^{-1}\right)^{kk'} \ a^{d'}_{l0} x_{j'}x_{k'}\nonumber\\
    =&  \langle (a^*)^d_{lm>0}x_jx_k\rangle \ \left(C^{-1}_l\right)^{dd'}\left(K^{-1}\right)^{jj'}\left(K^{-1}\right)^{kk'} \ a^{d'}_{lm>0} x_{j'}x_{k'} \nonumber\\
    &+\langle a^d_{lm>0}x_jx_k\rangle \ \left(C^{-1}_l\right)^{dd'}\left(K^{-1}\right)^{jj'}\left(K^{-1}\right)^{kk'} \ a^{*d'}_{lm>0} x_{j'}x_{k'} \nonumber\\
    & + \langle a^d_{l0}x_jx_k\rangle \ \left(C^{-1}_l\right)^{dd'}\left(K^{-1}\right)^{jj'}\left(K^{-1}\right)^{kk'} \ a^{d'}_{l0} x_{j'}x_{k'}.
\end{align}
If we repeat the process on all the variables, we obtain all the other combinations of complex conjugated variables. And now we can use the assumptions we have made about the isotropic bispectrum of the $a_{lm}$'s:
\begin{align}
\label{eq:app:7}
    \langle a^{d_1}_{l_1m_1}a^{d_2}_{l_2m_2}a^{d_3}_{l_3m_3} \rangle =&  \begin{pmatrix}
        l_1 & l_2 & l_3\\
        m_1 & m_2 & m_3
    \end{pmatrix} B_{l_1l_2l_3}^{d_1d_2d_3}.
\end{align}
Using the fact that $(a^d)^*_{lm}= (-1)^m a^d_{l-m}$  and that the $3j$-symbol in eq.~\eqref{eq:app:7} imposes $m_1+m_2+m_3=0$, we recover on the full space $m=-l,...,l$
\begin{align}
\label{eq:app:8}
    &\sum_{l_i,d_i,d'_i} \sum_{m_i}\begin{pmatrix}
        l_1 & l_2 & l_3 \\
        m_1 & m_2 & m_3
    \end{pmatrix} B_{l_1l_2l_3}^{d_1d_2d_3} \ 
    \left(C^{-1}_{l_1}\right)^{d_1d'_1} \left(C^{-1}_{l_2}\right)^{d_2d'_2} \left(C^{-1}_{l_3}\right)^{d_3d'_3} \ 
    \begin{pmatrix}
        l_1 & l_2 & l_3 \\
        -m_1 & -m_2 & -m_3
    \end{pmatrix} \widehat{B}_{l_1l_2l_3}^{d'_1d'_2d'_3}\nonumber\\
    &= \sum_{l_i,d_i,d'_i}B_{l_1l_2l_3}^{d_1d_2d_3} \ (-1)^{l_1+l_2+l_3}
    \left(C^{-1}_{l_1}\right)^{d_1d'_1} \left(C^{-1}_{l_2}\right)^{d_2d'_2} \left(C^{-1}_{l_3}\right)^{d_3d'_3} \ 
    \widehat{B}_{l_1l_2l_3}^{d'_1d'_2d'_3}.
\end{align}
For the linear part, we have terms such as
\begin{align}
\label{eq:app:9}
    &\langle x_ix_jx_k\rangle \ \left(K^{-1}\right)^{ii'}\left(K^{-1}\right)^{jk} \ x_{i'} \nonumber\\   
    &= \langle x_ix_jx_k\rangle \ \left(K^{-1}\right)^{ii'}\left(K^{-1}\right)^{jj'}\delta^{j'k} \ x_{i'} \nonumber\\
    &= \langle x_ix_jx_k\rangle \ \left(K^{-1}\right)^{ii'}\left(K^{-1}\right)^{jj'}(K \left(K^{-1}\right))^{j'k} \ x_{i'} \nonumber\\
    &= \langle x_ix_jx_k\rangle \ \left(K^{-1}\right)^{ii'}\left(K^{-1}\right)^{jj'} K^{j'k'} \left(K^{-1}\right)^{k'k} \ x_{i'} \nonumber\\
    &= \langle x_ix_jx_k\rangle \ \left(K^{-1}\right)^{ii'}\left(K^{-1}\right)^{jj'}  \left(K^{-1}\right)^{kk'} \ K^{j'k'} x_{i'} .\nonumber\\
\end{align}
Hence, we get the linear correction term as
\begin{align}
\label{eq:app:10}
    &\sum_{i,j,k,i',j',k'}-\langle x_ix_jx_k\rangle \ \left(K^{-1}\right)^{ii'}\left(K^{-1}\right)^{jj'}  \left(K^{-1}\right)^{kk'} \left[ K^{j'k'} x_{i'} + K^{i'k'} x_{j'} + K^{i'j'} x_{k'}\right]\nonumber\\
    &=\sum_{i,j,k,i',j',k'}-\langle x_ix_jx_k\rangle \ \left(K^{-1}\right)^{ii'}\left(K^{-1}\right)^{jj'}  \left(K^{-1}\right)^{kk'} \left[ \langle x_{j'}x_{k'}\rangle x_{i'} + \langle x_{i'}x_{k'}\rangle x_{j'} + \langle x_{i'}x_{j'}\rangle x_{k'}\right].
\end{align}
With reasoning similar to that used before, we recover
\begin{align}
\label{eq:app:11}
    =&\sum_{l_i,d_i,d'_i}-B_{l_1l_2l_3}^{d_1d_2d_3} \ \left(C^{-1}_{l_1}\right)^{d_1d'_1}\left(C^{-1}_{l_2}\right)^{d_2d'_2}  \left(C^{-1}_{l_3}\right)^{d_3d'_3}\sum_{m_i}\begin{pmatrix}
        l_1 & l_2 & l_3 \\
        m_1 & m_2 & m_3
    \end{pmatrix}\nonumber\\
    &\left[ \langle a^{d'_2}_{l_2-m_2}a^{d'_3}_{l_3-m_3}\rangle a^{d'_1}_{l_1-m_1} + \langle a^{d'_1}_{l_1-m_1}a^{d'_3}_{l_3-m_3}\rangle a^{d'_2}_{l_2-m_2} + \langle a^{d'_1}_{l_1-m_1}a^{d'_2}_{l_2-m_2}\rangle a^{d'_3}_{l_3-m_3}\right]\nonumber\\
    =&\sum_{l_i,d_i,d'_i}-B_{l_1l_2l_3}^{d_1d_2d_3} \ \left(C^{-1}_{l_1}\right)^{d_1d'_1}\left(C^{-1}_{l_2}\right)^{d_2d'_2}  \left(C^{-1}_{l_3}\right)^{d_3d'_3}(-1)^{l_1+l_2+l_3}\sum_{m_i}\begin{pmatrix}
        l_1 & l_2 & l_3 \\
        m_1 & m_2 & m_3
    \end{pmatrix}\nonumber\\
    &\left[ \langle a^{d'_2}_{l_2m_2}a^{d'_3}_{l_3m_3}\rangle a^{d'_1}_{l_1m_1} + \langle a^{d'_1}_{l_1m_1}a^{d'_3}_{l_3m_3}\rangle a^{d'_2}_{l_2m_2} + \langle a^{d'_1}_{l_1m_1}a^{d'_2}_{l_2m_2}\rangle a^{d'_3}_{l_3m_3}\right].
\end{align}
Finally, we get
\begin{align}
    \mathcal{L}_{NG}=&\Bigg\{1+\sum_{l_1\leq l_2\leq l_3} \frac{1}{g_{l_1l_2l_3}}\sum_{d_i,d'_i}B_{l_1l_2l_3}^{d_1d_2d_3} \ \left(C^{-1}_{l_1}\right)^{d_1d'_1}\left(C^{-1}_{l_2}\right)^{d_2d'_2}  \left(C^{-1}_{l_3}\right)^{d_3d'_3}(-1)^{l_1+l_2+l_3}\nonumber\\
    &\left[ \widehat{B}_{l_1l_2l_3}^{d'_1d'_2d'_3}-\sum_{m_i}\begin{pmatrix}
        l_1 & l_2 & l_3 \\
        m_1 & m_2 & m_3
    \end{pmatrix}\left( \langle a^{d'_2}_{l_2m_2}a^{d'_3}_{l_3m_3}\rangle a^{d'_1}_{l_1m_1} + \langle a^{d'_1}_{l_1m_1}a^{d'_3}_{l_3m_3}\rangle a^{d'_2}_{l_2m_2} + \langle a^{d'_1}_{l_1m_1}a^{d'_2}_{l_2m_2}\rangle a^{d'_3}_{l_3m_3} \right) \right]\Bigg\}\nonumber\\
    =&\Big( 1+ \langle B, \widehat{B} - B^{\mathrm{lin \ corr}}\rangle \Big),
\end{align}
since we can switch from the sum over all $l_i$ to the sum over the independent triplets $l_1\leq l_2\leq l_3$ by substituting the factor $6$ by $g_{l_1l_2l_3}$ previously defined for eq.~\eqref{eq:edge:5}.

\bibliographystyle{JHEP}   
\bibliography{refs}        

\end{document}